\documentclass{article}

\usepackage[english]{babel}

\usepackage[letterpaper,top=2cm,bottom=2cm,left=3cm,right=3cm,marginparwidth=1.75cm]{geometry}
\usepackage{csquotes}
\usepackage{amsmath}
\usepackage{algorithm}
\usepackage{dingbat}
\usepackage{algpseudocode}
\usepackage{graphicx}
\usepackage{soul}
\sethlcolor{yellow}
\usepackage[colorlinks=true, allcolors=blue]{hyperref}
\usepackage[
backend=biber,
style=ieee,
]{biblatex}
\usepackage{multirow}
\usepackage{booktabs}
\usepackage[table,xcdraw]{xcolor}
\usepackage{authblk}
\title{FedMicro-IDA: A Federated Learning and Microservices-based Framework for IoT Data Analytics}
\author[1]{Safa Ben Atitallah}
\author[*1,2]{Maha Driss}
\author[1]{Henda Ben Ghezela}

\affil[1]{RIADI Laboratory, University of Manouba, Manouba, Tunisia}
\affil[2]{Computer Science Department, CCIS, Prince Sultan University, Riyadh, Saudi Arabia}
\affil[*]{Corresponding author: mdriss@psu.edu.sa}

\DeclareUnicodeCharacter{0301}{\'{e}} 
\begin{document}
\maketitle

\begin{abstract}
The Internet of Things (IoT) has recently proliferated in both size and complexity. Using multi-source and heterogeneous IoT data aids in providing efficient data analytics for a variety of prevalent and crucial applications. To address the privacy and security concerns raised by analyzing IoT data locally or in the cloud, distributed data analytics techniques were proposed to collect and analyze data in edge or fog devices. In this context, federated learning has been recommended as an ideal distributed machine/deep learning-based technique for edge/fog computing environments. Additionally, the data analytics results are time-sensitive; they should be generated with minimal latency and high reliability. As a result, reusing efficient architectures validated through a high number of challenging test cases would be advantageous. The work proposed here presents a solution using a microservices-based architecture that allows an IoT application to be structured as a collection of fine-grained, loosely coupled, and reusable entities. The proposed solution uses the promising capabilities of federated learning to provide intelligent microservices that ensure efficient, flexible, and extensible data analytics. This solution aims to deliver cloud calculations to the edge to reduce latency and bandwidth congestion while protecting the privacy of exchanged data. The proposed approach was validated through an IoT-malware detection and classification use case. MaleVis, a publicly available dataset, was used in the experiments to analyze and validate the proposed approach. This dataset included more than 14,000 RGB-converted images, comprising 25 malware classes and one benign class. The results showed that our proposed approach outperformed existing state-of-the-art methods in terms of detection and classification performance, with a 99.24\% 


\end{abstract}

\section{Introduction}

Today, the Internet of Things (IoT) is a breakthrough technology that has changed how data are captured, gathered, and processed. It has improved the reality of the Internet by enabling the connection of an incredibly large number of disparate devices, including sensors and actuators \cite{hajjaji2021big}. These devices can interact and communicate with the real world, gather different types of environmental data, and support intelligent decision-making processes. IoT has made it simpler to design applications in a variety of fields, including smart agriculture, smart cities, and intelligent transportation systems, etc. \cite{atitallah2020leveraging, mezni2022smartwater,driss2022federated}. An IoT application is described as a collection of automated processes and data that are interconnected with other entities, including hardware and software. These entities communicate and share information with one another and the environment in order to accomplish their mutually agreed-upon goals.



Large volumes of data in several various formats are produced constantly by IoT entities. In order to support the operation of smart IoT applications, the acquired data is often transmitted to the cloud, where it is stored, processed, and analyzed. Employing data analytics on these huge data sets is essential to acquire significant insights, actionable knowledge, and relevant conclusions. Indeed, the use of data analytics techniques has become a prominent IT trend in academia, research, and business \cite{saleem2019data} \cite{atitallah2022microservices}. However, the IoT analytics community is confronted with new issues and challenges, including the processing complexity of the multisource IoT data, the treatment of different data types and formats, and the fast-growing data volumes. Such challenges negatively affect IoT application scalability, efficiency, and reusability. Moreover, IoT sensors and devices are typically located far from the central server where collected data are processed. Aside from being costly, transmitting these vast amounts of data that the edge devices have gathered to the central server exposes them to potential security risks \cite{nguyen2021federated}. 
Therefore, the centralized approach does not ensure the privacy of user data, and the cost is high in this approach. The main vulnerability in this approach is the central server; the entire system can fall offline, and operation might be stopped if the main server is hacked. By opting for  a decentralized and distributed approach, the system can continue to function even if one of the nodes or devices is hacked. Moreover, the distributed computing paradigm reduces latency by keeping data and moving processing resources closer to the end devices. In this context, Federated Learning (FL) has shown to be an effective approach for employing distributed data and computational resources to simultaneously train Machine Learning (ML) and Deep Learning (DL) models while ensuring end-user data privacy and security \cite{khan2021federated}. 
FL overcomes the limitations of centralized and edge computing paradigms, making them ideal data analytics tools for guaranteeing data privacy while exchanging information with peers. 

Recently, a novel learning paradigm has recently appeared and gained acceptance in a range of areas and applications, which is Transfer Learning (TL) \cite{atitallah2022novel,ben2022randomly,ben2022fusion}. The idea behind it is to employ previously learned models as the basis for subsequent challenges. This expedites training while simultaneously improving performance on related issues \cite{zhuang2020comprehensive}\cite{tan2018survey}. Moreover, TL utilizes the knowledge acquired from the core huge dataset to assist in solving new issues and challenges faced in other case studies dealing with modest amounts of data \cite{zhuang2020comprehensive}\cite{tan2018survey}. Recent studies have shown that combining the TL approach with FL produces more fruitful results for a variety of case studies \cite{shi2022deep}\cite{kevin2021federated}\cite{he2020group}.
To reuse the capabilities offered by TL and FL approaches,  the adoption of a microservices-based design is highly advantageous \cite{al2022ai}. The microservices architecture is based on the idea of building a large-scale software application out of a collection of loosely coupled services \cite{champaneria2023microservices}\cite{atitallah2022microservices}\cite{driss2021microservices}. 
A microservices-based architecture is inherent to the distributed nature of IoT ecosystems. Numerous benefits are delivered by using a microservices-based architecture for IoT applications. The fact that a system's services are mostly decoupled from one another makes it simple to build, modify, and scale the entire IoT application. Furthermore, reusability, dependability, and agility are key advantages of the microservices-based design. Since microservices are self-contained and autonomous software units, this fact offers independently deployable services that can be reused, extended, or discarded without changing the business logic of the entire application \cite{surianarayanan2019essentials}.

This work presents a novel data analytical solution based on a microservices-oriented design that uses FL and TL methodologies to leverage and gain insights from IoT ecosystems. The proposed solution, named FedMicro-IDA, consists of several software components designed to support distributed learning and satisfy the needs of IoT applications. Relying on the FL and TL methodologies, we advocate an intelligent microservices-based approach ensuring proper data privacy and security and providing effective analytics to end users. We suggest breaking the IoT application into a linkable collection of microservices that run across multiple computing layers (i.e., cloud and edge layers). We focus on providing an adequate service computing distribution through the application of the FL approach that ensures data secrecy and reduces transmission costs. The proposed solution enables the efficient, secure, and adjustable distribution of data processing at edge devices, which can aid in developing sustainable IoT data analytical implementations.


The principal contributions of this paper are outlined in the following points:
\begin{itemize}
    \item We proposed an intelligent microservices-based framework to support distributed data analytics by employing the FL and TL paradigms for the context of IoT applications. The essential components of data analytics were defined, developed, and deployed as microservices. We combined Web semantic technologies with AI techniques to create intelligent, straightforward, and composable microservices that guaranteed effective IoT data analytics;
    
    \item A malware detection and classification case study was used to validate the proposed approach. Data privacy was protected by the application of the FL paradigm, which did not need data transfer to the cloud and allowed training to be carried out locally. Furthermore, the TL methodology's prior knowledge improved the proposed analytical model, boosting its accuracy as a result.
    
    \item For the experiments, we used a  public dataset, MaleVis, which contains over 14,000 RGB images from 26 unique families. It includes 25 classes of malware of five distinct categories (adware, Trojan, Virus, Worm, Backdoor) and one benign class.
    
    \item To evaluate the suggested approach properly, We conducted a thorough performance analysis utilizing several assessment metrics, including accuracy, precision, recall, F1-score, and loss. We also compared our approach with the centralized model. We developed the distributed FL and the centralized variant of our proposed approach using the TensorFlow DL framework to compare them. We also assess how well the approach's built-in microservices perform. Microservices' execution times are examined to assess the effectiveness of the microservices-based architecture.
    
\end{itemize}
The present article is divided into six sections. Section 2 presents the microservices-based architecture, TL, and FL methodologies. Section 3 examines recent relevant works about DL/ML/FL-based data analytics in IoT environments. In Section 4, the proposed approach is presented and detailed in depth. Section 5 summarizes the experiments and analyzes the findings. Finally, Section 6 highlights the major contributions made by this study and suggests future research directions.

\section{Background}
This section covers three major topics: microservices-based architecture, TL, and FL. The microservices design is perfect for assuring low latency and high dependability, as well as the overall scalability and availability of the IoT system. TL and FL are two innovative machine/deep learning approaches that have successfully supplied intelligence to many distributed applications and platforms.

\subsection{Microservices-Based Architecture}
Most of the problems with identifying, categorizing, and describing services and composition solutions have been resolved by the adoption of the conventional service-based architecture. However, the internet of things/services brought up new problems. The ecosystem in which services are designed, implemented, and executed has become more open, dynamic, and changeable. This brings up a number of concerns with malleability, such as the capacity of self-configuring, self-optimizing, self-healing, and self-adapting services. This may entail devices with constrained resources and processing capabilities, necessitating the development of innovative techniques for dynamically managing such lightweight and basic services \cite{bucchiarone2020microservices}.
Additionally, context awareness, heterogeneity, device dependencies, and personalization must be taken into account while managing services in today's ubiquitous contexts. In order to guarantee the interoperability of diverse things, such as mobile devices, sensors, and networks, a ubiquitous environment necessitates adequate semantic technologies, common standards, and mediation \cite{surianarayanan2019essentials}. In this context, microservices emerged as the best technical solution to deal with these issues at the organizational scale by ensuring the aggregation of several services while maintaining a good quality of service (QoS) and minimizing coupling levels.
With IoT data analytics, microservices-based architecture provides faster and more efficient results. The processing time is speeding up, and the average latency is reduced. The adoption of microservices-based architecture for IoT applications provides numerous advantages \cite{atitallah2022microservices}\cite{driss2021microservices}\cite{hasan2021sublmume}.
The key benefits of using a microservices-based architecture are stated in the list below: \cite{driss2022ws}\cite{driss2022req}:
\begin{itemize}
    \item Independent development: all microservices can be easily developed based on their individual functionality;
    \item Independent deployment: based on their services, they can be individually deployed in any application;
    \item Fault isolation: even if one service of the application does not work, the system continues to function;
    \item Mixed technology stack: different languages and technologies can be used to build different services of the same application;
    \item Granular scaling: individual components can scale as per need; there is no need to scale all components together
\end{itemize}
In addition to the advantages already mentioned, the creative architecture built on microservices may be improved by the use of artificial intelligence (AI) technologies to ensure the development of new intelligent IoT apps that offer sophisticated AI data analytics as microservices. Our primary goal in this paper is to accomplish this by combining containerization technologies with DL and ML models.

\subsection{Transfer Learning}

The main idea of the TL approach is to use the gained knowledge to address new challenges and problems \cite{zhuang2020comprehensive}. Instead of building and training DL models from scratch, which takes a long time, a huge number of computational resources, and a significant volume of data, leveraging pre-trained Convolutional Neural Networks (CNN) architectures by reusing existing layer weights will speed up and enhance the learning process. 

TL approach has lately garnered significant appeal across a wide range of industries and research topics, notably in IoT applications, in which it was adopted to improve the ability to create value-added applications from cutting-edge and effective AI-based data analytical methods \cite{atitallah2022novel,wu2020exploiting,ohata2020automatic}. Its main goal is to reuse previously learned models as starting points for tackling new problems and tasks. This significantly increases performance on related problems (i.e., compared to creating models from scratch approach) while reducing training time. Image classification is one of the most well-known and explored applications of DL and TL. The principle behind TL for image classification is that if a model is trained on a large and varied enough dataset, it may serve as a generic model of the visual world.
TL has recently been applied to federated settings and has considerably increased the effectiveness, and the generalization ability of different frameworks dealing with various issues related to IoT ecosystems \cite{chen2020fedhealth,liu2020secure,zhang2022federated}.  
It permitted minimizing the computation and communication costs of the federation systems while improving the overall performance of the learning algorithms.

In the field of TL, several CNN architectures can be found. MobileNetV2, DenseNet201, and InceptionV3 are examples that are described in depth in the following subsections:

\subsubsection{MobileNetV2}
The MobileNetV2 architecture \cite{sandler2018mobilenetv2} is an improved version of the first launched MobileNet CNN model. This novel design is based on an inverted residual structure, which is made up of bottleneck layers that include convolutional blocks. A skip connection technique is used for each convolutional block's starting and ending points. Using a thin skip connection strategy, the MobileNetV2 can retrieve prior activations that had not been altered for each convolutional block. In terms of performance, MobileNetV2 outperforms previous versions and is computationally more efficient.
    
\subsubsection{DenseNet201}
Recently, Huang et al. \cite{huang2017densely} proposed a very deep CNN model with a feed-forward structure for connecting layers, named the DenseNet201 model. Each layer in this model takes its inputs from the extracted feature maps of the previous layer, and its output extracted feature maps are used as input for the successive layers. This technique significantly reduced the number of parameters, improved performance with repeated feature maps, and alleviated vanishing gradient difficulties. However, because this model has many layers, training takes much longer than in other DL models.

\subsubsection{InceptionV3}
The InceptionV3 model \cite{szegedy2015going} is a deep network that consists of 48 layers. It is a reworking version of the original Inception model with greater depth and processing effectiveness. Several convolutional layers with varying filter sizes (i.e., (1×1), (3×3), and (5×5)) operate simultaneously in the InceptionV3 model. The model can capture local and extended features across small and large convolutions because of its hybrid filter architecture. Because of its parallel computation, this model outperforms other existing CNN architectures in several case studies.

\subsection{Federated Learning}

Recently, Google researchers presented the FL paradigm as a promising approach for addressing the bandwidth requirements, privacy protection, and regulatory issues encountered in the internet of services/things ecosystems, pervasive computing, and mobile applications  \cite{abreha2022federated}.
An FL approach follows the centralized distributed learning strategy in which models are trained on end devices without sharing their local datasets. The training of models is entailed on separate and distant devices where the data remains locally. FL algorithms have enabled the creation of a wide range of IoT applications, including monitoring and surveillance systems, smart grids, and cyberattack detection solutions, while maintaining confidentiality and privacy by storing training data locally, adapting to potential changes, and significantly improving training results \cite{shaik2022fedstack,su2021secure,driss2022federated}. In this context, FL is an intriguing paradigm that has shown remarkable success in delivering intelligence to the IoT edge layers. Participating devices use local data to train a global model started by a central authority autonomously \cite{li2020review}. It comes with various benefits:
\begin{itemize}
    \item FL allows edge devices to learn from predictive models and preserve training datasets instead of hosting them on a centralized server.
    \item The data is accessible on edge devices, making real-time AI model changes possible. This feature saves time and makes data available without requiring a connection to the centralized server.
    \item FL ensures data security management by storing the data locally on the edge device.
    \item FL algorithms are appropriate for deployment on resource-constrained devices due to their low complexity and distributed structure.
\end{itemize}
The FL training process is broken down into five phases \cite{driss2022federated}. First, the FL server chooses an ML model to be trained locally on each client node. Second, a subset of current client nodes is chosen at random or using appropriate selection algorithms. Third, the server sends the initial global model to the chosen client nodes. Clients download the current global parameters of the model and train it locally. Each client node communicates updates to the server during the fourth phase. Finally, without gaining access to the client's data, the FL server receives the updates and aggregates them using aggregation algorithms to generate a new global model. The FL server orchestrates the training process and communicates the global model modifications to the chosen client at the end of each cycle. These phases are repeated until the required level of quality performance is achieved.

\section{Related Work}

Developing IoT applications while ensuring interoperability, scalability, and dependability raises multiple issues. These issues are principally caused by the ultra-large scale of Internet of Things ecosystems, the heterogeneous nature of apps and devices, and the tremendously dynamic nature of the execution environment. Several research studies have examined the development of IoT applications by opting for a microservices-based architecture on numerous use cases to overcome these issues. The following paragraphs present recent and relevant microservices-based approaches designed to enhance data analytics in IoT ecosystems.  

In \cite{jamil2021intelligent}, Jamil et al. proposed an intelligent and secure microservice system based on blockchain technology to enhance the predictive modeling of fitness data in an IoT scenario. The suggested system supports microservice-based analytic features to deliver safe and dependable IoT services. The collected data's privacy, security, and identity are ensured with a permissioned blockchain network. Besides, fitness users' customized workout and diet plans are prepared using the predictive analytics microservice. This recommendation model is built on a collection of microservices, including data pre-processing, feature extraction, model selection, and visualization. This design helps to provide a good performance and deliver accurate results, ensuring the scalability and reliability of the fitness application.

In \cite{vresk2016architecture}, Vresk et al. provided a suggestion for an architecture of a microservices-based middleware aimed at connecting heterogeneous IoT devices. The proposed approach seeks to mask the complexity caused by end-device property variation by using a consistent method to model both physical and logical IoT devices and services. As a validation use case, the estimation of future energy usage is employed. A simple linear regression model is designed to depict the correlation between an outcome and one or more predictors. The result is a trained regression model that can be used to score new samples to make predictions.

Dineva et al. \cite{dineva2020architectural} presented a microservices-based architectural framework that provides ML solutions as services. The IoT data analytics are enabled through a set of microservices, each offering a single analytical function. Identity management, data acquisition, data transformation, ML, device management, and logging are the functions provided as microservices in the proposed architecture.

Ortiz et al. \cite{ortiz2019real} implemented a microservice-based architecture for predictive data processing. The architecture is supplied with compelling predictions through predictive data analytics, which allows taking real-time actions. The implementation of microservices-based architecture has substantially enhanced the suggested framework's maintenance and scalability. An air quality management case study was used to assess the framework's performance.

In \cite{ali2018design}, Ali et al. proposed a microservices-based methodology that provides efficient data analytical modules implemented as autonomous microservices. They divided the data analytics process into a set of major steps and developed them as microservices. To enhance the application's scalability, the suggested modules were deployed as microservices, composed, and orchestrated with each other. The proposed design was implemented by a prototype and evaluated using an IoT use case scenario. support vector machine (SVM), linear discriminant analysis (LDA), k-nearest neighbors (K-NN), decision trees (DT), and linear regression (LR) were the ML models used to analyze and learn from data. The experimental results demonstrated the enhancement of the IoT applications' performance in terms of robustness, scalability, and stability.

In \cite{abdel2021federated}, Abdel-Basset et al. introduced a Fed-TH model, based on DL and FL, for detecting cyber attacks in industrial cyber-physical systems. This model was deployed as an edge microservice, with an exploratory microservice handling the latency issue. The Fed-TH model was validated using two public datasets, and it provided an accuracy of 92.97\% and 92.84\% for the TON-IOT and LITNET-2020 datasets, respectively.

By adopting both TL and FL methodologies, Chen et al. \cite{chen2020fedhealth} proposed the FedHealth platform used for healthcare support. FedHealth created the DL models using TL and trained them by adopting FL settings. The final developed models provided precise and customized healthcare programs for users while maintaining their privacy and security. 
%

In \cite{houmani2021enabling}, Houmani et al. proposed an architecture that facilitated the DL workflows, reduced the implementation time, and enhanced the analytics results. The proposed approach was validated with an object detection use case by employing edge and cloud resources. The contributions made by this work were as follows: 1) a data management approach that emphasized data quality adaptability by controlling the trade-offs between latency and accuracy and changing the resolution of the data sources and 2) a workflow scheduling method based on data that allocated DL tasks over the computer spectrum.

In \cite{roy2021micro}, Roy et al. proposed an architecture based on microservices and DL named Micro-safe to provide safety as a service in a 6G environment. This architecture produced customized safety-related decisions and sent them to the registered end-users using a Deep Neural Network (DNN). The experimental results demonstrated the efficiency of the suggested architecture, which delivered better latency, lower energy usage, and higher throughput.

A customizable framework based on microservices and visualization technologies has been proposed in \cite{nikolakis2020microservice} to support intelligent predictive analytics and maintenance operations in industrial IoT contexts. The design of this platform was structured as a set of microservices, where every service performed a specific function. The proposed platform enabled scalable data storage and management. Moreover, the predictive facilities were provided as a service to facilitate the use of trained models considered for analytics.

The study in \cite{attota2021ensemble} presented a decentralized FL architecture named MV-FLID. This architecture was designed to offer an intrusion detection solution. An analytical ML model was constructed and trained on various perspectives of IoT network data to recognize, classify, and prevent intrusions. In addition, the proposed model was based on multi-view ensemble learning to enhance overall learning efficiency. 

In \cite{bibi2022deep}, the authors offer a self-learning independent multi-vector threat analysis and detecting methodology to safeguard IIoT systems effectively. The Cuda-powered Convolutional LSTM2D mechanism, which they developed, is modular and self-optimizing, and can effectively deal with dynamical versions of upcoming IIoT risks and assaults. The suggested approach is tested on a cutting-edge dataset with 21 million examples of distinct attack patterns and threat vectors. A comparison with modern DL-driven architectures and benchmark algorithms shows that the suggested methodology delivers greater accuracy in threat detection.

According to the investigations conducted in the previously detailed related works, microservices-based architectures for IoT data analytics are becoming more intriguing, investigated, and implemented. Despite this, given the numerous existing learning methodologies/algorithms and schemas to choose from, more study is needed to determine the convenient solution for IoT ecosystems.

We may conclude the following  deficiencies from our review of the previously presented relevant related works:

\begin{itemize}
    \item The level of privacy and security is compromised since the entire dataset must be sent to the cloud when using the centralized learning approach to run analytics in the cloud;
    \item Most of the reported works used conventional ML algorithms. However, these algorithms are unable to work effectively with unstructured data. This requires hand-crafted feature engineering, which takes time and effort;
    \item Some of the previous studies using DL models are time-consuming and need considerable computational resources to ensure required data analytics tasks. To attain high-performance outcomes, the employed DL models need to be trained throughout a high number of epochs;
    \item The majority of the research relied on a single data analytics technique to carry out analytical tasks, omitting the advantages of combining many techniques, which might significantly enhance performance outcomes compared to those provided by applying a single technique.
    
\end{itemize}

To address the limitations found in the state-of-the-art approaches previously discussed, our proposed solution provides the following key benefits:
\begin{itemize}
    \item Decompose the IoT application into a loosely coupled set of microservices that execute on various compute tiers (cloud/edge). The learning tasks will be accomplished by distributed computing microservices and will be efficiently and easily maintained, expanded, and reused;
    \item By utilizing the TL approach, we aim to improve learning and give end users relevant data analytics. The key advantage of this type of learning is that the model training phase can be managed with limited resources and in less time compared to other commonly used data analytics models. A model that has already been trained on a task with plenty of labeled training data will handle a different task  with much fewer data;
    \item Use the FL approach to boost analytics performance while protecting the privacy and security of collected IoT data. This approach addresses the limitations of both centralized and edge computing paradigms by providing an effective data analytics tool by offering a powerful and effective data analytics tool that ensures data privacy while exchanging knowledge with peers;
    \item Utilize the ensemble learning approach to aggregate results from several DL classifiers and optimize the performance of the analytics task;
    \item  Design, implement, and validate an all-in-one framework, FedMicro-IDA, offering effective, extensible, and secure analytical capabilities that serve for the examination of different IoT-based case studies and the elaboration of comprehensive performance analysis by considering various significant assessment measures.
    
\end{itemize}


\begin{table}[h]
\caption{Summary of the reviewed research works}
\label{tab:RW}
\resizebox{\textwidth}{!}{%
\begin{tabular}{|c|c|c|c|c|c|c|c|c|c|}
\hline
\textbf{Work} &
  \textbf{\begin{tabular}[c]{@{}c@{}}Application\\ domain\end{tabular}} &
  \textbf{\begin{tabular}[c]{@{}c@{}}Learning\\ scheme\end{tabular}}  &
  \textbf{\begin{tabular}[c]{@{}c@{}}Data\\ analytics\end{tabular}} &
  \textbf{\begin{tabular}[c]{@{}c@{}}Analytical\\ task\end{tabular}} &
  \textbf{\begin{tabular}[c]{@{}c@{}}Data\\analytics\\ technique(s)\end{tabular}} &
  \textbf{\begin{tabular}[c]{@{}c@{}}Service\\$/$Monolitic\\ architecture\end{tabular}}&
  \textbf{\begin{tabular}[c]{@{}c@{}}Use of\\FL\end{tabular}}&
  \textbf{\begin{tabular}[c]{@{}c@{}}Use of\\TL\end{tabular}}&
  \textbf{\begin{tabular}[c]{@{}c@{}}Learning \\fusion\end{tabular}}
  \\ \hline
{\cite{jamil2021intelligent}} &
  Healthcare &
  \begin{tabular}[c]{@{}c@{}}Centralized\\ computing\end{tabular}  &
  \begin{tabular}[c]{@{}c@{}}Prescriptive \\ analytics\end{tabular} &
  \begin{tabular}[c]{@{}c@{}}Recommendation \\ of diet and\\ workout plan\end{tabular} &
  \begin{tabular}[c]{@{}c@{}}SVM,KNN,\\ LR\end{tabular} 
  &\begin{tabular}[c]{@{}c@{}}Blockchain-\\based \\ microservice\\ \checkmark\end{tabular}
 
  & \textbf{X}
  & \textbf{X}
  & \textbf{X} \\ \hline
  {\cite{vresk2016architecture}} &
  \begin{tabular}[c]{@{}c@{}} Smart\\ environment\end{tabular}&
  \begin{tabular}[c]{@{}c@{}}centralized\\ computing\end{tabular}&
  \begin{tabular}[c]{@{}c@{}}Descriptive\\ analytics\end{tabular} &
  \begin{tabular}[c]{@{}c@{}}  Estimation of \\future energy\\ consumption \end{tabular} &
  \begin{tabular}[c]{@{}c@{}}Regression\\ model\end{tabular} 
  &  \begin{tabular}[c]{@{}c@{}}Microservices- \\based \\ architecture \\ \checkmark \end{tabular} 
  & \textbf{X}
  & \textbf{X}
  & \textbf{X}
  \\ \hline
  
{\cite{dineva2020architectural}} &
  \begin{tabular}[c]{@{}c@{}}High abstract\\ conceptual\\ approach\end{tabular} &
  \begin{tabular}[c]{@{}c@{}}Centralized\\ computing\end{tabular}   &
  \begin{tabular}[c]{@{}c@{}}Predictive \\ analytics\end{tabular} &
  Prediction &
  \begin{tabular}[c]{@{}c@{}}The different\\  techniques of\\  ML\end{tabular} 
  &  \begin{tabular}[c]{@{}c@{}}Microservices\\ with ASP.NET \\ \checkmark \end{tabular}
  & \textbf{X}
  & \textbf{X}
  & \textbf{X}
  \\ \hline
{\cite{ortiz2019real}} &
  Air quality &
  \begin{tabular}[c]{@{}c@{}}Centralized \\ computing\end{tabular} &
  \begin{tabular}[c]{@{}c@{}}Predictive\\  analytics\end{tabular} &
  \begin{tabular}[c]{@{}c@{}}Event pattern \\ detection, \\ real-time prediction, \\ context awareness\end{tabular} &
  \begin{tabular}[c]{@{}c@{}}ARIMA \\ model\end{tabular} 
    & \begin{tabular}[c]{@{}c@{}}REST \\ services \\ \checkmark \end{tabular}
  & \textbf{X}
  & \textbf{X}
  & \textbf{X}
  \\ \hline
{\cite{ali2018design}} &
  Healthcare &
  \begin{tabular}[c]{@{}c@{}}Centtralized \\ computing\end{tabular}&
  \begin{tabular}[c]{@{}c@{}}Predictive\\ analytics\end{tabular} &
  \begin{tabular}[c]{@{}c@{}}Classification of\\ positive or \\ negative diabetes \\ patients\end{tabular} &
  \begin{tabular}[c]{@{}c@{}}SVM, \\ KNN, DT,\\  LR\end{tabular} 
  & \begin{tabular}[c]{@{}c@{}}Microservices- \\based \\ architecture \\ \checkmark \end{tabular}
  & \textbf{X}
  & \textbf{X}
  & \textbf{X}
   
  \\ \hline
{\cite{abdel2021federated}} &
  \begin{tabular}[c]{@{}c@{}}Industrial \\ threat hunting\end{tabular} &
  \begin{tabular}[c]{@{}c@{}}Distributed \\ computing\end{tabular} &
  \begin{tabular}[c]{@{}c@{}}Predictive \\ analytics\end{tabular} &
  \begin{tabular}[c]{@{}c@{}}Cyber-threat\\  detection\end{tabular} &
  LSTM-AE 
  &  \begin{tabular}[c]{@{}c@{}}Microservices- \\based \\ architecture \\ \checkmark \end{tabular}
  & \checkmark
  & \textbf{X}
  & \textbf{X}

  \\ \hline

   {\cite{houmani2021enabling}}
   &  \begin{tabular}[c]{@{}c@{}}Smart \\ city\end{tabular}
   &  \begin{tabular}[c]{@{}c@{}}Distributed\\ Computing\\ (cloud+\\fog+edge)\end{tabular}
   &  \begin{tabular}[c]{@{}c@{}}Predictive\\ analytics\end{tabular}
   &  \begin{tabular}[c]{@{}c@{}}Object detection\\ in images and \\ videos\end{tabular}
   &  \begin{tabular}[c]{@{}c@{}}YOLO \\ algorithm\end{tabular}
   & \begin{tabular}[c]{@{}c@{}}Microservices- \\based \\ architecture \\ \checkmark \end{tabular}
   & \textbf{X}
   & \checkmark
   & \textbf{X}
  
   \\ \hline 
   \cite{roy2021micro}
   &  \begin{tabular}[c]{@{}c@{}}Smart \\transportation\end{tabular}
   & \begin{tabular}[c]{@{}c@{}}Distributed\\ computing\end{tabular}
   
   & \begin{tabular}[c]{@{}c@{}}Predictive \\analytics\end{tabular}
   & \begin{tabular}[c]{@{}c@{}} Address the safety\\ problems in\\ transportation services. \end{tabular}
   & \begin{tabular}[c]{@{}c@{}}DNN \end{tabular}
  & \begin{tabular}[c]{@{}c@{}}Microservices- \\based \\ architecture \\ \checkmark \end{tabular}
  & \textbf{X}
  & \textbf{X}
  & \textbf{X}
  
  \\ \hline
     {\cite{nikolakis2020microservice}} &
  \begin{tabular}[c]{@{}c@{}}Smart\\ manufacturing\end{tabular} &
  \begin{tabular}[c]{@{}c@{}}Distributed\\ computing\\ (edge)\end{tabular} &
  \begin{tabular}[c]{@{}c@{}}Predictive\\ analytics\end{tabular} &
  \begin{tabular}[c]{@{}c@{}}Support smart\\ predictive \\ maintenance\\  operations\end{tabular} &
  \begin{tabular}[c]{@{}c@{}}Pretrained \\ analytical\\ models\end{tabular} 
  &\begin{tabular}[c]{@{}c@{}}Microservices- \\based \\ architecture \\ \checkmark \end{tabular}
  & \textbf{X}
  & \textbf{X}
   & \checkmark
   
  \\ \hline
  {\cite{chen2020fedhealth}} &
  Healthcare &
  \begin{tabular}[c]{@{}c@{}}Distributed \\ computing\end{tabular} &
  \begin{tabular}[c]{@{}c@{}}Predictive \\ analytics\end{tabular} &
  \begin{tabular}[c]{@{}c@{}}Human activity \\ recognition\end{tabular} &
  CNN with TL 
  & \begin{tabular}[c]{@{}c@{}}Monolitic\\ architecture \\ \textbf{X} \end{tabular}
  & \checkmark
  & \checkmark
  & \textbf{X}
  
  \\ \hline
    {\cite{attota2021ensemble}} &
  Cyber security &
  \begin{tabular}[c]{@{}c@{}}Distributed \\ computing\end{tabular} &
  \begin{tabular}[c]{@{}c@{}}Predictive \\ analytics\end{tabular} &
  \begin{tabular}[c]{@{}c@{}}Intrusion\\ detection\end{tabular} &
  ML models
  &\begin{tabular}[c]{@{}c@{}}Monolitic\\ architecture \\ \textbf{X} \end{tabular}
  & \checkmark
  & \textbf{X}
  & \checkmark
  
  \\ \hline
\begin{tabular}[c]{@{}c@{}}Proposed \\ FedMicro-IDA \end{tabular} &
  \begin{tabular}[c]{@{}c@{}}Generic \\solution\end{tabular} &
  \begin{tabular}[c]{@{}c@{}}Distributed\\ computing  \\ (FL)\end{tabular} &
  \begin{tabular}[c]{@{}c@{}}Descriptive$/$\\ Diagnostic$/$\\Predictive$/$\\Prescriptive\\ analytics\end{tabular} &
  \begin{tabular}[c]{@{}c@{}}Any type of \\analytical task\end{tabular} &
  \begin{tabular}[c]{@{}c@{}}Pretrained\\ CNN (TL)\end{tabular}
  &\checkmark & \checkmark
  & \checkmark
  & \checkmark
  
  \\ \hline
\end{tabular}}
\end{table}
%
%
\section{Proposed Approach}
 
This work focuses on developing an intelligent approach that supports IoT data analytics with distributed learning. This approach is implemented through a set of microservices to be secure, flexible, scalable, and able to react faster to user demands.
In fact, the data analytics functionality is split into discrete containerized microservices, and the analytics models are trained using the FL method and deployed on edge computing nodes.
\\

The proposed  framework is named as  \textbf{Fed}erated Learning and \textbf{Micro}services-based Framework for \textbf{I}oT \textbf{D}ata \textbf{A}nalytics (FedMicro-IDA). FedMicro-IDA has been abstracted into two 
main layers: the cloud layer 
and the edge layer. 
IoT devices play a pivotal role in the process of data collection in this suggested approach. These IoT devices are purpose-built and equipped to gather data from various sources efficiently. Strategically positioned near the client's edge, they enable swift and effective data acquisition. 
The data preprocessing stage is performed on edge devices, notably edge servers. Each client has its own dedicated edge server, which oversees the entire analytical process, including preprocessing and training. We have enabled localized data cleaning, transformation, and feature extraction before starting the training stage by incorporating data preprocessing at the edge level.

\begin{figure}[h]
\centering
\includegraphics[scale=0.6]{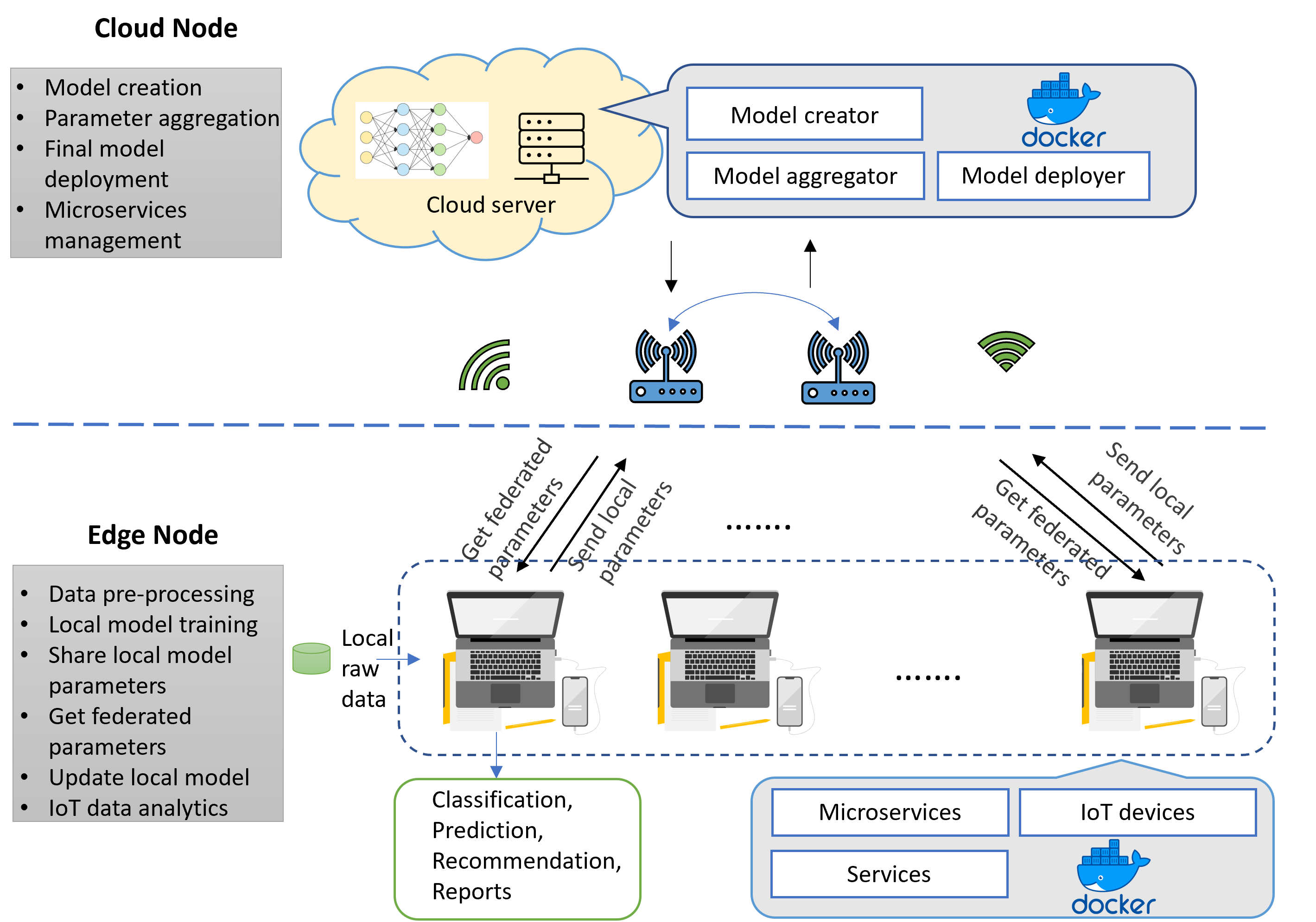}
\caption{Abstract architecture of FedMicro-IDA}
\label{fig:pict1}
\end{figure}

Figure \ref{fig:pict1} illustrates the abstract architecture design of FedMicro-IDA. We summarized the approach functionalities and highlighted the responsibilities of each component in the following subsections. 
%
\subsection{Cloud Node: FL server components}
The FL server components are developed and encapsulated into a set of microservices. Each microservice delivers a specific function. 
The construction of the training model is the initial stage in the FL process. A set of microservices in the central server are responsible for model creation, initialization, training configuration, evaluation, and deployment. These microservices are:
\begin{itemize}
    \item {Model creator:} responsible for creating the initial global model and its compilation;
    \item {Model aggregator:} based on the supplied local models from the clients, a new global model is aggregated and delivered as an output;
    \item  {Model deployer:} following the aggregation step, the resulting global model's performance is examined and sent to the client devices for data analytics.
\end{itemize}
%
\subsection{Edge Node: FL client components} \label{sec2.4}
\begin{figure}[h]
\centering
\includegraphics[scale=0.6]{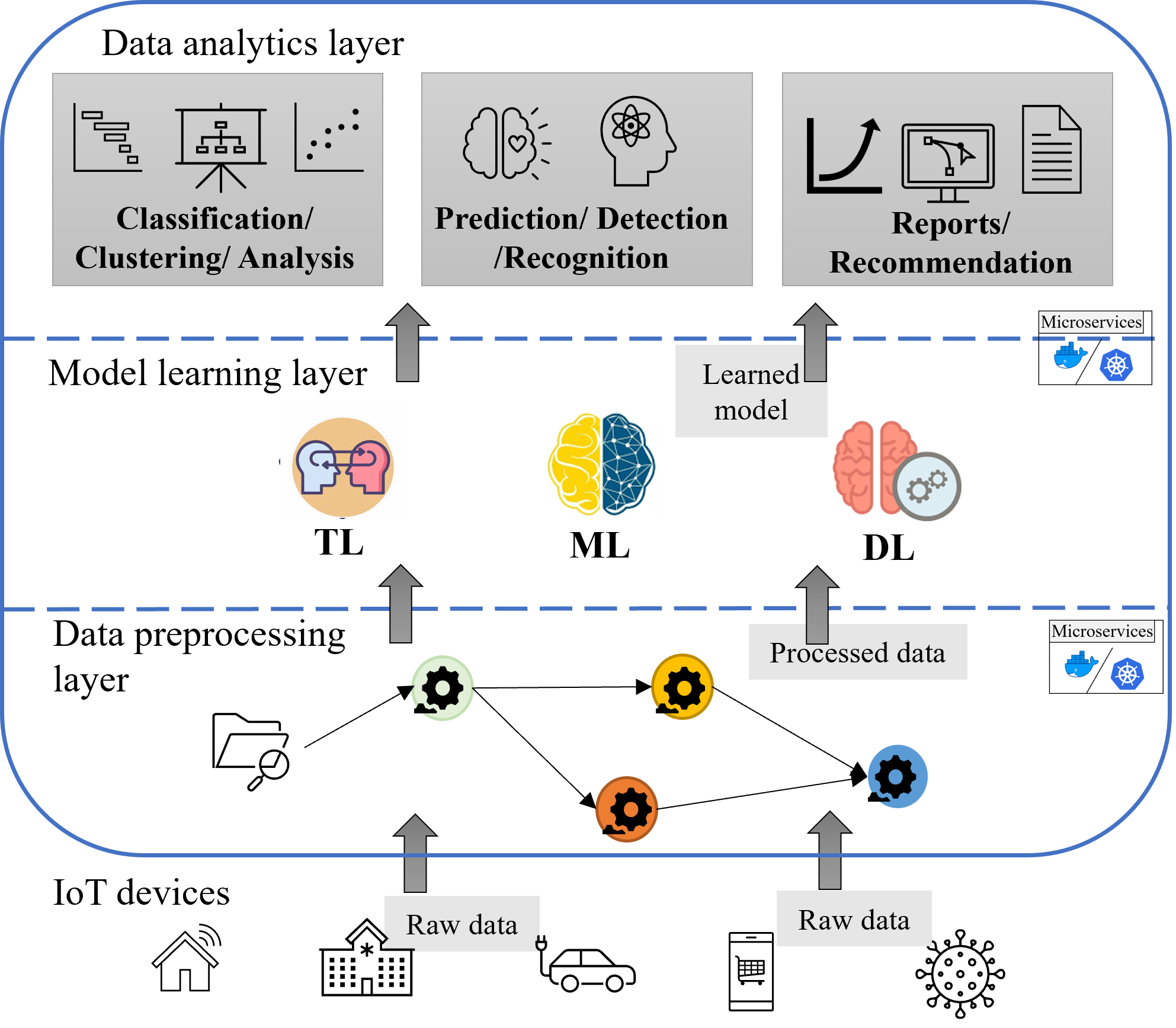}
\caption{FL client's components.}
\label{fig:edge}
\end{figure}

The edge layer consists in a layered service-oriented system that runs based on linkable microservices to form an intelligent IoT application, as illustrated in Figure \ref{fig:edge}. These layers are described in the following points: 
\begin{enumerate}
    \item \textbf{ Data preprocessing:\\} 
    
    To enable the execution of various data operations, the edge server takes charge of processing and managing the raw data collected from IoT devices. This involves employing the service-oriented computing paradigm, which allows for the creation of a suite of services specifically designed to handle distinct data management and preprocessing tasks tailored to each specific use case. By adopting this approach, the edge server acts as a facilitator, streamlining the processing of IoT data and enabling the seamless execution of a range of operations on the data.
    Three microservices have been developed to be applied to the collected data.:
    \begin{itemize}
        \item {Data cleaning:} it defines the data filtering process, where the noise is eliminated, the missing values are resolved, and the unnecessary data is removed;
       \item {Data integration:} since the data is gathered from different sources, it should be integrated and structured with the same format;
       \item {Data scaling/transformation:} the gathered data is normalized and transformed to a defined range in order to enhance the consistency of the data.
    \end{itemize}
    This client-side preprocessing strategy has several benefits. First, minimizing the requirement to transmit raw data to a central processing unit can effectively decrease communication overhead. Instead, the edge devices preprocess the data, thus lowering the amount of data that needs to be sent. Additionally, this approach improves privacy by restricting sensitive or raw data communication to higher-level systems by only sending processed and pertinent data.
   
    \item \textbf{ Model learning:\\}
    After the creation in the central server, the client-server receives the developed model and starts the learning process. This stage consists of three microservices, each of which is responsible for a predefined function as the following:
    \begin{itemize}
        \item {Model training:} using the prepared data, the model starts the training based on the configuration parameters specified by the global model;
        \item {Model evaluator:} The performance of the local model is measured in the local model evaluator to be then uploaded to the model aggregator in the federated server if the performance requirement is met;
        \item {Model uploader:} The model uploader is responsible for transferring the trained local model parameters to the server to ensure the aggregation process. 
    \end{itemize}
   
    \item \textbf{Data analytics:\\} The resulting global model is used to assess new data, then produces the desired insights according to the case study under consideration (e.g., classification, clustering, prediction, reports, recommendations, etc.).
    \begin{itemize}
        \item Analytics microservice: prepares the model and uses the new data to make the appropriate analytics. Visualization microservices utilize the obtained analytic findings to build more particular views and get data insights; 
        \item Results fusion: the fusion method is used to merge data analysis findings from different global models. Learning fusion aims to enhance individual classifier decisions by producing more thorough and accurate decisions.
    \end{itemize}
    
\end{enumerate}

\subsection{Learning Process}
The FedMicro-IDA framework opts for a Federated Transfer Learning (FTL) of different CNN models to increase and enhance the learning process. The training of CNN models is carried through a number of virtual instances on its own training data at the edge. The updated parameters are exchanged with the central server in the cloud for aggregation purposes. Once the global models are developed, they are used on edge to provide required data analytics.

\begin{figure}[h]
\centering
\includegraphics[scale=0.6]{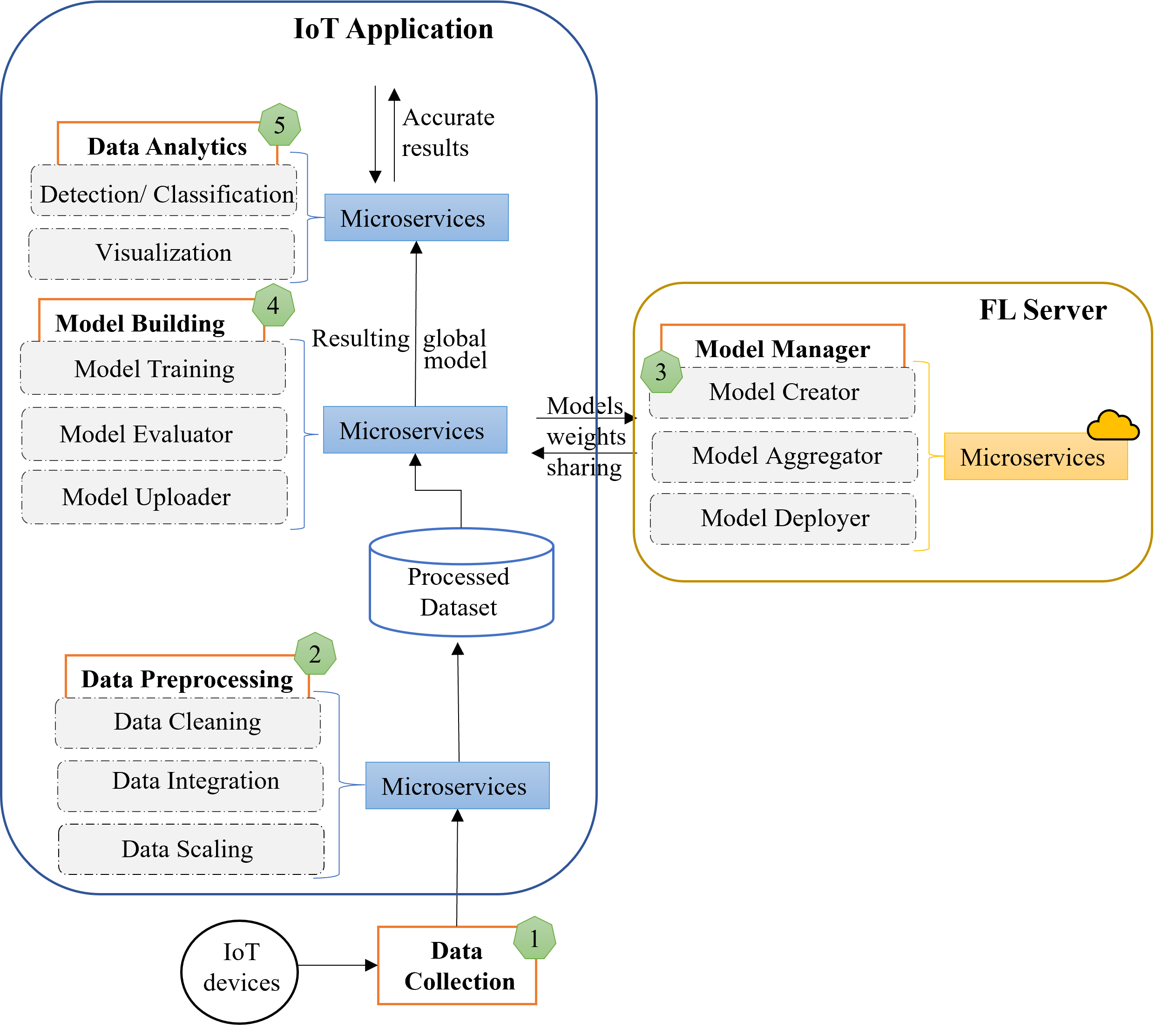}
\caption{Learning process proposed in FedMicro-IDA.}
\label{fig:pict2}
\end{figure}

The detailed learning process flow between the FL server and the client-server is shown in Figure \ref{fig:pict2}.
The working procedure of the proposed FL method is summarized in the following points:
\begin{enumerate}
    \item In each client, data is collected by IoT devices and prepared to be trained locally;
    \item The initial DL models are configured in the cloud server and shared with the edge client servers;
    \item After acquiring  the DL models, the training process is performed with the client's local data throughout multiple epochs;
    \item The new weights of local models are uploaded to the server for FL aggregation; 
    \item The weights of DL models are aggregated in the FL server, and new global DL models are constructed. The FL training process (Steps 3 and 4) continues for predefined communication rounds until the models reach effective performance results;
    \item The new global model weights are transferred back to FL clients. The existing local models in the clients are replaced with the resulting global ones;
    \item The analytics task(s) is/are performed on the client's local data; 
    \item A fusion ensembler is developed to receive the prediction probabilities from the resulting global models and select the final effective outputs.
\end{enumerate}

The resulting global model is specifically defined as the model obtained at the end of the FL training. Although it is labeled as "resulting", it is important to recognize that the FL process involves iterative updates and refinements. These updates can be influenced by various factors, such as the incorporation of additional data, integration of feedback mechanisms, and utilization of optimization techniques. Hence, it is essential to acknowledge that the resulting global model represents a significant milestone in the FL process, while also considering the ongoing potential for iterative improvements and refinements based on the aforementioned factors.

The aggregation method for locally trained DL models in client servers is conducted as many times as the number of communication rounds in the FL process.
By exchanging just trained models' parameters rather than local data, this strategy decreases the device's communication overhead and ensures privacy.
In each training cycle, the local updates provided by each client in FL are merged using an aggregation function. FedAvg \cite{mcmahan2017communication} represents the most fundamental aggregation function, generating the global model based on the average of the weights given by the FL clients. Suppose $W=(w_{i})$ is the weights of the generated model and $W^k = (w_{i}^k)$ the weights of each client $k$, then:
\begin{equation}
w_{i} = \sum \frac{d_{i}}{D} w_{i}^k  ,
\end{equation}
where $D$ is the total data size and and $d_{i}$ is the data size of each client.
Technically, the aggregation process is done according to equations (\ref{eq:1}) and (\ref{eq:2})
\begin{equation}\label{eq:1}
    F_{k}(w) = \frac{1}{d_{k}} \sum_{i \in P_{k}} f_{i}(w) ,
\end{equation}

\begin{equation}\label{eq:2}
    f(w)=\sum_{k=1}^{K} \frac{d_{k}}{D} F_{k}(w) .
\end{equation}

Where the total number of included clients is denoted as $K$, the loss of prediction of a sample  $(x_{i}; y_{i})$  is denoted as $f_{i}(w)$ = $l(x_{i}$; $y_{i} w)$, and $P_{k}$ is a collection of data sample indices for client k.
Equation \ref{eq:1} determines the weight parameters of all devices based on the loss values experienced across the training data points.
Equation \ref{eq:2} scales the parameters and adds them all up component-by-component.

 
After training and obtaining the resulting global models, we used the Dempster–Shafer fusion theory to merge the extracted probabilities acquired by the DL global models \cite{sentz2002combination,denoeux2000neural,ben2022fusion}. This evidence theory is based on three crucial functions: the mass ($m$),  the belief ($Bel$), and the plausibility ($Pl$).

The mass function represents the basic probability assignment, which is based on evidence theory and described in the following Equations \ref{eq:eq3}, \ref{eq:eq4}, and \ref{eq:eq5}:
\begin{equation}
\label{eq:eq3}
    2^D \rightarrow [0.1], 
\end{equation}
\begin{equation}
\label{eq:eq4}
     m(\O)=0,
\end{equation}
\begin{equation}
\label{eq:eq5}
     \sum_{A \subset D}m(A)=1,
\end{equation}

$\forall$A $\in 2^D$, $m(A)$ is the mass function of the proposition that represents the basic belief degree and the initial support degree strictly assigned to proposition $A$. $D =
(P_{1},P_{2},…,P_{n}), P_{i}(1\leq i \leq n)$ are the sets of probabilities predictions outputted from the resulting global models.

The Bel belief function is a totally increasing function on $2^D \rightarrow $[0.1], such as Bel($\O$)=0  and Bel($D$)=1. Bel functions are defined according to Equations \ref{eq:bel} and \ref{eq:bel2}.
\begin{equation}
    \label{eq:bel}
    \forall A \in 2^D, 
\end{equation}
\begin{equation}
    \label{eq:bel2}
   Bel(A) =\sum_{B \subseteq A, B \neq \O } m(B)
\end{equation}

The plausibility functions are set to $Pl(A) =1-Bel(A)$, where $Pl(A)$, and $Bel(A)$ are the upper and lower limit functions of $A$, respectively. To combine probabilities coming from the different architectures of CNN models, the orthogonal rule of the Demspher fusion is used as depicted in Equation \ref{eq:mass7}.
\begin{equation}
    \label{eq:mass7}
    m(A)=(m_{1}\oplus m_{2} \oplus ...\oplus m_{l})(A)= \frac{\sum_{B_{1}\cap...\cap B_{l}}m_{1} (B_{1}) m_2(B_{2}).. m_{l}(B_{l}) }{ 1-K}
\end{equation}

where
\begin{equation}
    \label{eq:k}
    K = \sum_{B_{1}\cap ... \cap B_{l}=\O} m_{1}(B_{1})m_{2}(B_{2})...m_{l}(B_l),
\end{equation}

$K$ represents the degree of conflict between the $l$ different architectures of DL.
In this study, the rule used for decision-making is the maximum of belief presented in Equation \ref{eq:x}.
\begin{equation}
    \label{eq:x}
    x \in C_{i},  if Bel(C_{i})(x)=max [Bel(C_{k})(x), 1\leq k \leq n ]. 
\end{equation}

The whole process of the proposed fusion-based FTL approach is summarized in Algorithm \ref{algo:FL}
\begin{algorithm}[ht]
\caption{: Fusion-based FTL process}
\label{algo:FL}
\begin{algorithmic}[1]
\State \textbf{Input:}  N different datasets {$DS_{1}$, $DS_{2}$, ..,  $DS_{n}$}.
\State \textbf{Output:} Prediction/detection .
\State \textbf{Begin}
\State  Set the models’ initial parameters for all the clients
\State $mw_{1}$, $mw_{2}$, $mw_{3}$  \textcolor{blue}{/* The local DL models’ weights */}

\State  C= c1, c2, …, cn       \textcolor{blue}{  /* Initializing the FL clients  */}
\State Reading input data
\Function FL training (CommunicationRrounds): 
    \While{ $c_{i}$ in C do:}
       \State \textcolor{blue}{ /*  train the models with local data of each client  */}
       \State  $mw_{1}$ = train( data in $c_{i}$) 
       \State  $mw_{2}$ = train( data in $c_{i})$  
       \State  $mw_{3}$ = train( data in $c_{i}$) 
       \State  return $mw_{1}$, $mw_{2}$, $mw_{3}$
   \EndWhile
\EndFunction
\Function FLaggregation:
\State \textcolor{blue}{/* Averaging parameters of each model from all devices using Eq(\ref{eq:1}) and(\ref{eq:2})*/}
\State $Mw_1$ = FLaverage($c_1(mw_1)$, $c_2(mw_1)$, .., $c_n(mw_1)$) 
\State $Mw_2$ = FLaverage($c_1(mw_2)$, $c_2(mw_2)$, .., $c_n(mw_2)$) 
\State $Mw_3$ = FLaverage($c_1(mw_3)$, $c_2(mw_3)$, .., $c_n(mw_3)$) 
\State Replace local models with global models 
\State return $Mw_{1}$, $Mw_{2}$, $Mw_{3}$  
\EndFunction
\Function  Ensembler ($Mw_{1}$, $Mw_{2}$, $Mw_{3}$) :
 \For{\texttt{each client}}
    \State newdata  \textcolor{blue}{/*to be used for data analytics*/}   
    \State predictions[] = $Mw_{1}$(newdata), $Mw_{2}$(newdata), $Mw_{3}$(newdata) 
    \State final prediction = ensemble(predictions)
 \EndFor
\EndFunction
\end{algorithmic}
\end{algorithm}

\section{Implementation and  Evaluation of FedMicro-IDA}
 
This section presents the use case scenario considered for the proposed framework's validation, the dataset utilized for experiments, the environment setup, and descriptive details about the simulation experiments. In the last subsection, we present and analyze the experimental results.
\subsection{Use Case Scenario: IoT Malware Detection and Classification}
\begin{figure}[]
\centering
\includegraphics[scale=0.60]{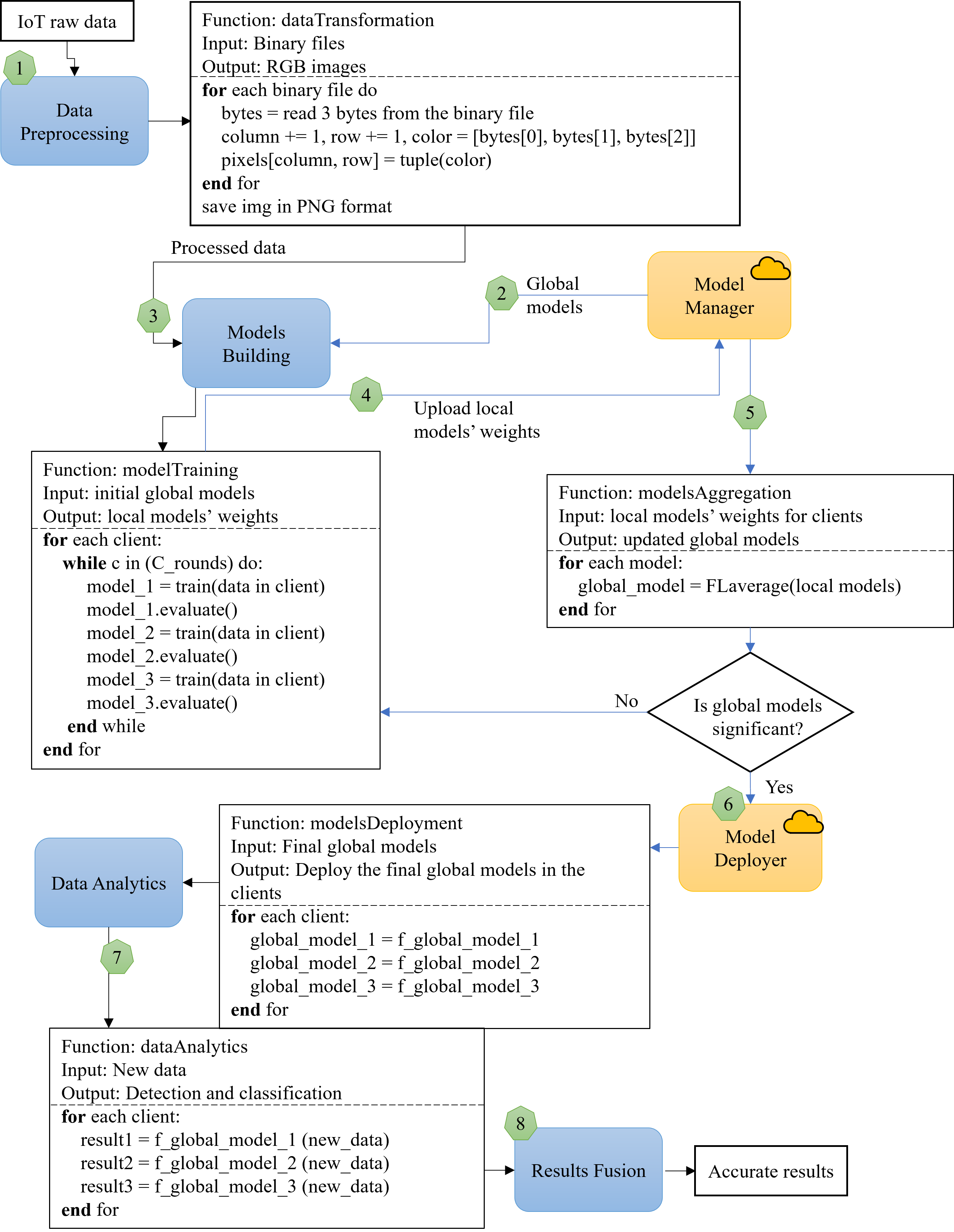}
\caption{FedMicro-IDA for IoT malware detection and classification Scenario.} 
\label{fig:pict44}
\end{figure}

We implement an IoT-malware detection and classification use case to validate and assess the proposed approach. Figure \ref{fig:pict44} represents in detail the different stages of the use case scenario. The high-level elements of FedMicro-IDA include IoT devices, malware detection entities (i.e., clients and servers), and FL servers. The data traffic is accumulated, preprocessed, and analyzed for each client using the proposed malware detection solution provided by FedMicro-IDA. This solution consists of microservices that support data preprocessing, model learning, and data analytics. The flow of data analytics is carried out with an FL server to detect and classify various malware effectively. Each microservice encapsulates a distinct set of capabilities, such as data transformation, model training, and model uploader.
 
\begin{figure}[]
\centering
\includegraphics[width=\textwidth]{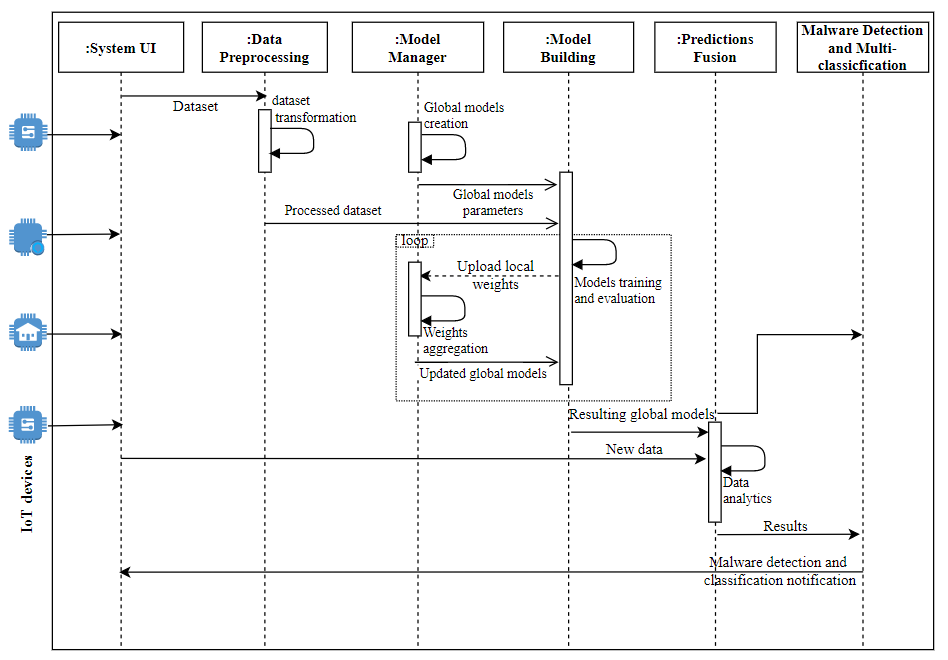}
\caption{FedMicro-IDA in operation for IoT malware detection and classification.}
\label{fig:pict5}
\end{figure}

The primary purpose of the considered use case is to develop an effective solution that provides accurate IoT malware detection and classification results. Combining the FL approach with TL allows for the delivery of an efficient model trained with data collected from various environments while maintaining the privacy and security of this data. Adopting a microservices-based architecture improves the suggested solution's extensibility, reuse, reliability, and security. Figure \ref{fig:pict5} illustrates the malware detection and classification scenario using a sequence diagram. This scenario is initiated by various microservices collecting IoT traffic data. This data is gathered and saved as binary files that must be processed and converted into images. The generated images are transmitted to the TL CNN, which is trained using the FL methodology. The resulting models are used to assess new data and provide insights once the CNNs have been built and trained. In the last phase, the obtained results are combined using the Demspher theory to offer effective detection and classification results.

\subsection{Dataset} 
For the evaluation and validation of our proposed approach, we used the malware evaluation with vision (MaleVis) dataset \cite{MaleVis}. This dataset contains 14,226 RGB-converted photos from 26 distinct IoT malware families. It covers 25 types of malware divided into five classes: adware, trojans, viruses, worms, backdoors, and one benign class. Table \ref{tab:table2} presents a brief overview of each of these malware classes.
For experiments, the dataset was partitioned into 70\%, 20\%, and 10\% for training, validation, and testing, respectively.

\begin{table}[ht]
\centering
\caption{Malware classes in Malevis. }
\label{tab:table2}
\begin{tabular}{|c|c|c|c|}
\hline
\begin{tabular}[c]{@{}l@{}}\textbf{Malware}\\ \textbf{category}\end{tabular} 
 &
  \multicolumn{1}{c|}{\textbf{Description}} &
  \textbf{Classes} &
  \textbf{Samples} \\ \hline
Adware &
  \begin{tabular}[c]{@{}c@{}} Malicious application that uses your web browser\\ to show advertisements on your screen \end{tabular} &
  \begin{tabular}[c]{@{}c@{}}Adposhel, Amonetize,   \\ BrowseFox, InstallCore, \\ MultiPlug, Neoreklami\end{tabular} &
  2983 \\ \hline
Trojan &
  \begin{tabular}[c]{@{}c@{}}Malicious code that is designed to damage,\\disrupt, steal, or in general \\negatively affect your data or network\end{tabular} &

  \begin{tabular}[c]{@{}c@{}}Agent ,Dinwod, Elex, \\ HackKMS, Injector,\\  Regrun, Snarasite, \\ VBKrypt, Vilsel\end{tabular} &
  4440 \\ \hline
Virus &
  \begin{tabular}[c]{@{}c@{}}A malware that duplicates itself by \\inserting code into other programs\end{tabular} &
  \begin{tabular}[c]{@{}c@{}}Neshta, Sality,\\ Expiro, VBA\end{tabular} &
  1974 \\ \hline
Worm &
  \begin{tabular}[c]{@{}c@{}} An isolated piece of malicious programs\\that propagates to other systems \\over a computer network\end{tabular} &
  \begin{tabular}[c]{@{}c@{}}Allaple, Autorun, \\ Fasong, Hlux\end{tabular} &
  19974 \\ \hline
Backdoor &
  \begin{tabular}[c]{@{}c@{}}Malware that circumvents common \\ authentication mechanisms to access a system\end{tabular} &
  Androm, Stantinko &
  1000 \\ \hline
\end{tabular}
\end{table}

\begin{figure}[h]
\centering
\includegraphics[width=\textwidth]{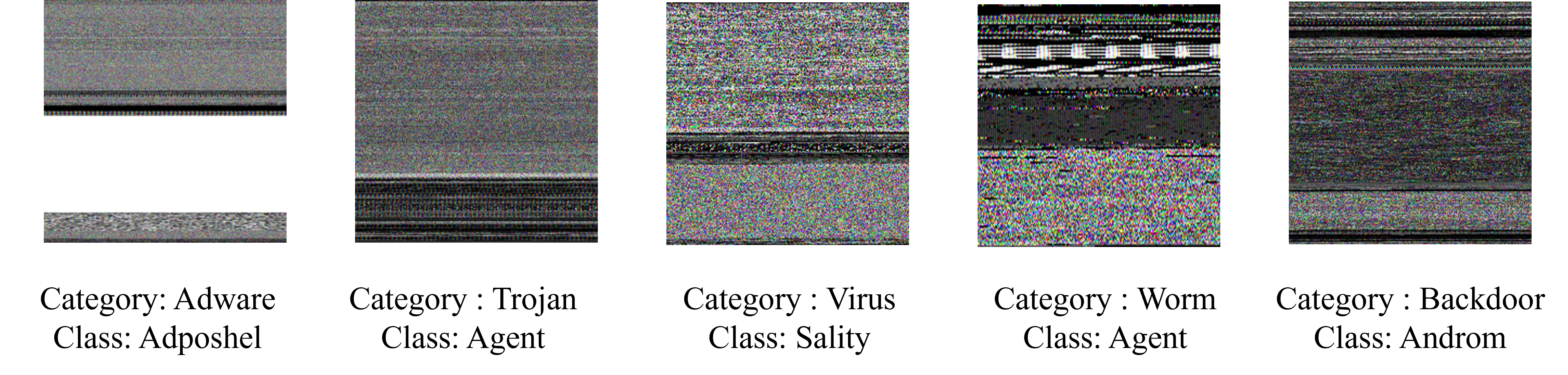}
\caption{Samples of malware images in Malevis.}
\label{fig:pict_m}
\end{figure}

Figure \ref{fig:pict_m} illustrates some samples of malware images found in the MaleVis dataset for adware, trojan, virus, worm, and backdoor classes.

\subsection{Environment Setup} 
FedMicro-IDA for IoT malware detection and classification was implemented using a PC with the following characteristics: an Intel(R) Core(TM) i7-8565U CPU @ 1.80 GHz 1.99 GHz processor, 16 GB RAM, and an NVIDIA GeForce MX graphics card running on Windows 11. The developed microservices were deployed with Docker images \cite{docker}. The deployed Docker images formed a cluster of containers that were orchestrated using swarm orchestration \cite{kubernetes}.
The DL models were implemented using Python 3.8 \cite{Python} and developed within the Jupyter notebook environment \cite{Jupyter} provided by the Anaconda package \cite{Anaconda}. In our implementation, we leveraged both the Keras library \cite{keras}, along with the TensorFlow backend \cite{tensorflow}. For the development and deployment of the FL methodology, we utilized TensorFlow Federated (TFF) \cite{ftensorflow}, an open-source framework specifically designed for decentralized ML and DL computations. TFF enabled us to experiment and explore the capabilities of FL.

\subsection{Evaluation Metrics} 
In order to assess the performance of the proposed approach, accuracy, precision, recall, and F1-score metrics were used.
The statistical measures are represented mathematically in Equations \ref{eq:4} – \ref{eq:10}, where:
\begin{itemize}
    \item True Positive (TP): Malware (i.e., positive sample) is the predicted result, and the prediction is correct;
    \item True Negative (TN): Benign class (i.e., negative sample) is the predicted result, Normal, and the prediction is correct;
    \item False Positive (FP): Malware is the predicted result, and the prediction is incorrect;
    \item False Negative (FN): Benign class is the predicted result, and the prediction is incorrect;
\end{itemize}
\textbf{Accuracy:} It is used to evaluate the model's overall performance throughout all categories. 
\begin{equation}\label{eq:4}
    Accuracy=  \frac{TP+TN}{TP+TN+FP+FN}
\end{equation}
\textbf{Precision:} It is used to assess the model's accuracy in classifying a sample as positive or negative. 
\begin{equation}\label{eq:5}
    Precision=  \frac{TP}{TP+FP}
\end{equation}
\textbf{Recall :} it is employed to assess the model's ability to identify the positive samples. 
\begin{equation}\label{eq:6}
    Recall=  \frac{TP}{TP+FN}
\end{equation}
\textbf{F1-score:} It combines the accuracy and recall measurements to produce a value-added rating for performance verification.
\begin{equation}\label{eq:8}
    F1-score=  \frac{2*Precision*Recall}{Precision+Recall}
\end{equation}
\textbf{Specificity:} It evaluates the model's ability to detect negative samples.
\begin{equation}\label{eq:10}
    Specificity=  \frac{TN}{TP+FN}
\end{equation}
\textbf{ Matthews Correlation Coefficient (MCC):} It is calculated by multiplying the correlation coefficient between the actual and predicted classes.
\begin{equation}\label{eq:9}
    MCC= \frac{(TP*TN)-(FP*FN)}{\sqrt{(TP+FP)(TP+FN)(TN+FP)(TN+FN)}}  
\end{equation}
\textbf{Loss:} It is used to estimating the error value and assess how effectively the model handles data.\\

\subsection{Details on Simulation Experiments}

\begin{figure}[h]
\centering
\includegraphics[scale=0.9]{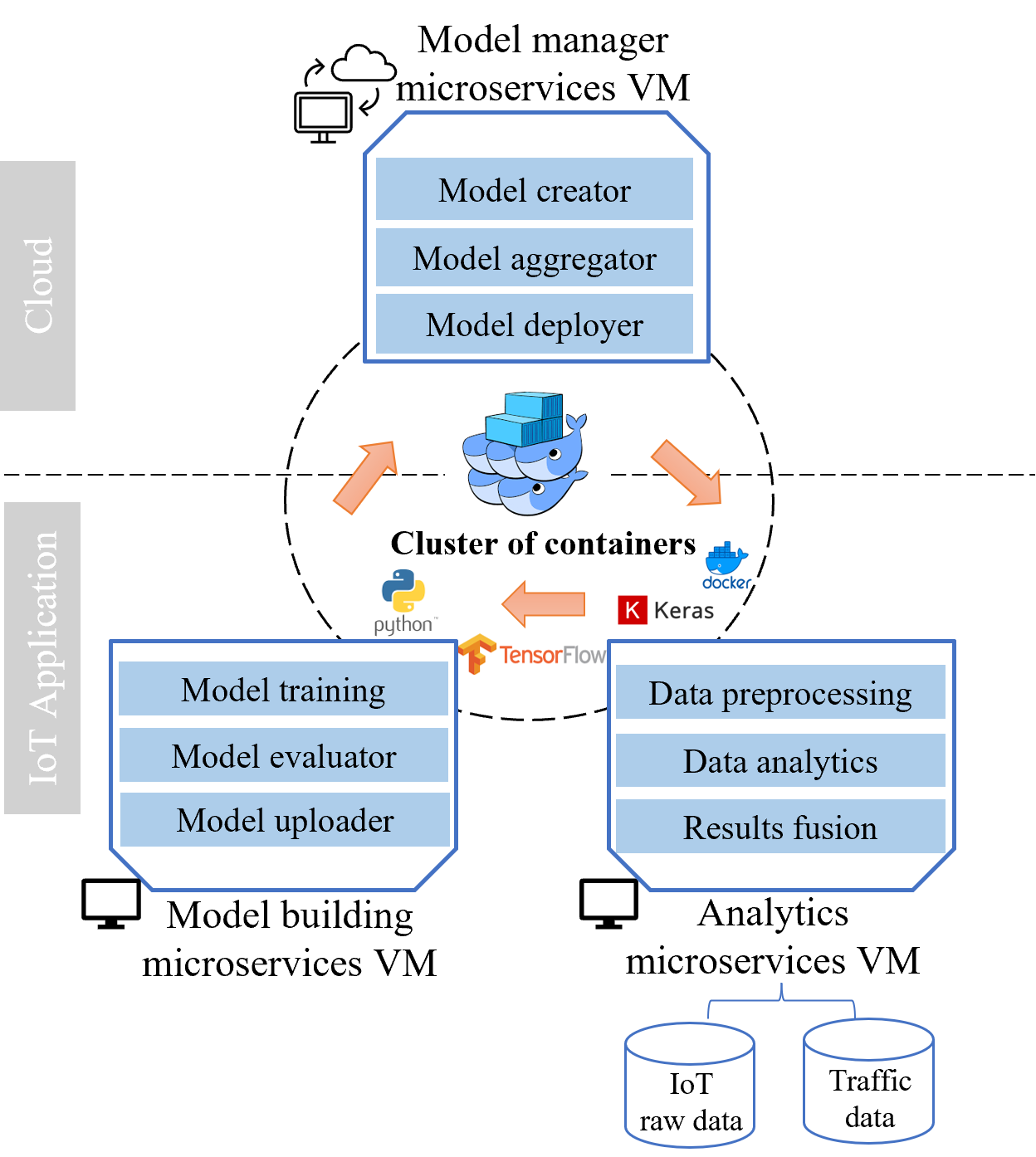}
\caption{Experimental setup of FedMicro-IDA used for the IoT malware detection and classification use case scenario.}
\label{fig:pict_d}
\end{figure}

Figure \ref{fig:pict_d} shows simulation experiments designed to accomplish the primary goal of FedMicro-IDA, which is the adoption of microservices and AI techniques to ensure predictive analytics in IoT environments. FedMicro-IDA was designed and validated using open-source components that were publicly accessible. Adopting a microservices-based architectural style and FL/TL approaches assured the suggested solution's scalability and reliability. It is also worth noting that microservices have been successfully integrated and used with FL and TL methodologies.

With distributed docker images, three virtual machines have been constructed \cite{docker}. A microservice was hosted by each Docker instance installed on the virtual machine. The Docker images that were installed formed a cluster of containers. Swarm orchestration \cite{smith2017docker} was used to orchestrate the components of this cluster. The high degree of availability given for apps is one of the key benefits of running a Docker swarm. The first virtual machine was designed to support the model manager processes, which made use of the model creator, aggregator, and deployer microservices. The second virtual machine was set up for model-learning processes that comprise microservices for models' training, evaluation, and uploading. The last virtual machine was in charge of the data analytics algorithms that were implemented using data preprocessing and results' fusion microservices.

\subsection{Experimental Results}
This subsection reports and analyses the results of applying the FedMicro-IDA approach to the IoT-malware detection and classification use case using the MaleVis dataset.
\subsubsection{Analytics Results 
}
\begin{figure}[h]
\centering
\includegraphics[scale=0.4]{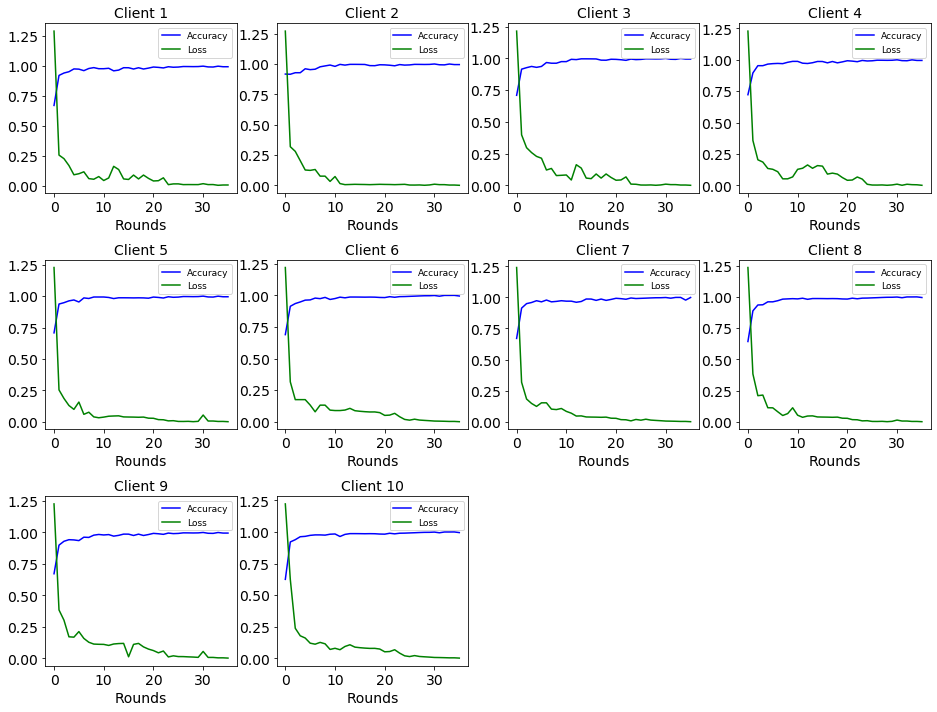}
\caption{Accuracy and loss curves of the MobileNetV2 model trained with adversary clients over 35 communication rounds.}
\label{fig:pict6}
\end{figure}

To evaluate the proposed approach, it was implemented using the TFF framework \cite{tff} by considering 10 virtual client servers with a federated setting. Three pre-trained CNN architectures, including MobileNetV2, DensNet201, and InceptionV3, were used with the TL methodology to detect and efficiently classify IoT malware. 35 communication rounds have been considered for models' aggregation, where the CNNs are trained over 10 epochs in each round using its local data. We used the Adam optimizer with a learning rate of 1e-3 and the cross-entropy loss function to configure the DL models. The size of the input images was 224x224 pixels, while the batch size was 32. 
To enable the self-learning process, the dataset was distributed into different sets of training and test data over 10 different clients.
\\Figure \ref{fig:pict6} illustrates the accuracy and loss curves of the MobileNetV2 model after each aggregation round. This model has achieved a high training accuracy, near 100\%, at round 20 for all the clients. 

To conduct a comparative analysis, we also implemented our proposed approach's Centralized Learning model (CL) using the TensorFlow DL framework \cite{tf}. The MobileNetV2, DenseNet201, and InceptionV3 were fine-tuned and trained on the entire MaleVis dataset. In contrast to the FL implementation, the knowledge produced by the centralized approach is shared in the cloud, and the amount of data utilized for training is enormous compared to the amount of data used in the FL approach. The CNNs were trained over 35 epochs. 
Comparing the CNN models' performance results emphasizes that the FL implementation outperforms the CL implementation. 

\begin{figure}[h]
\centering
\includegraphics[scale=0.5]{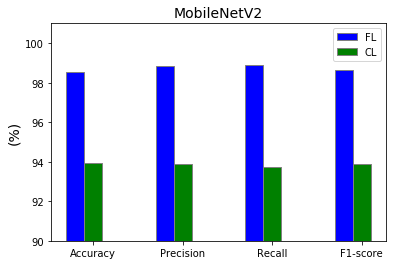}
\caption{Performance results of the MobileNet model resulting from the application of the FL and CL approaches.}
\label{fig:pict8}
\end{figure}

Table \ref{tab:tab4} summarizes the performance results of the fine-tuned CNN models trained using the CL and FL methodologies in addition to the fusion-based classifiers of these models. The results presented by Table \ref{tab:tab4} indicate that the performance of models trained in FL is better than those trained in CL approach. It is clear from Figure \ref{fig:pict8} that the accuracy, precision, recall, and F1-score metrics of the MobileNetV2 model, which provided the best performance results by applying the FL methodology, surpass the performance results of the same model by using the CL approach.

After the training, Demspher's theory was applied to this scenario. The performance results of the generated classifiers are presented in Table \ref{tab:tab4}. In terms of accuracy, the combined classifiers clearly outperform the individual models. The proposed FedMicro-IDA achieves better performance with 99.24\% accuracy, 99.12\% precision, 99.36 recall, and 99.45\% F1-score.  

%
%
 
\begin{figure}[!h]
\centering
\includegraphics[scale=0.35]{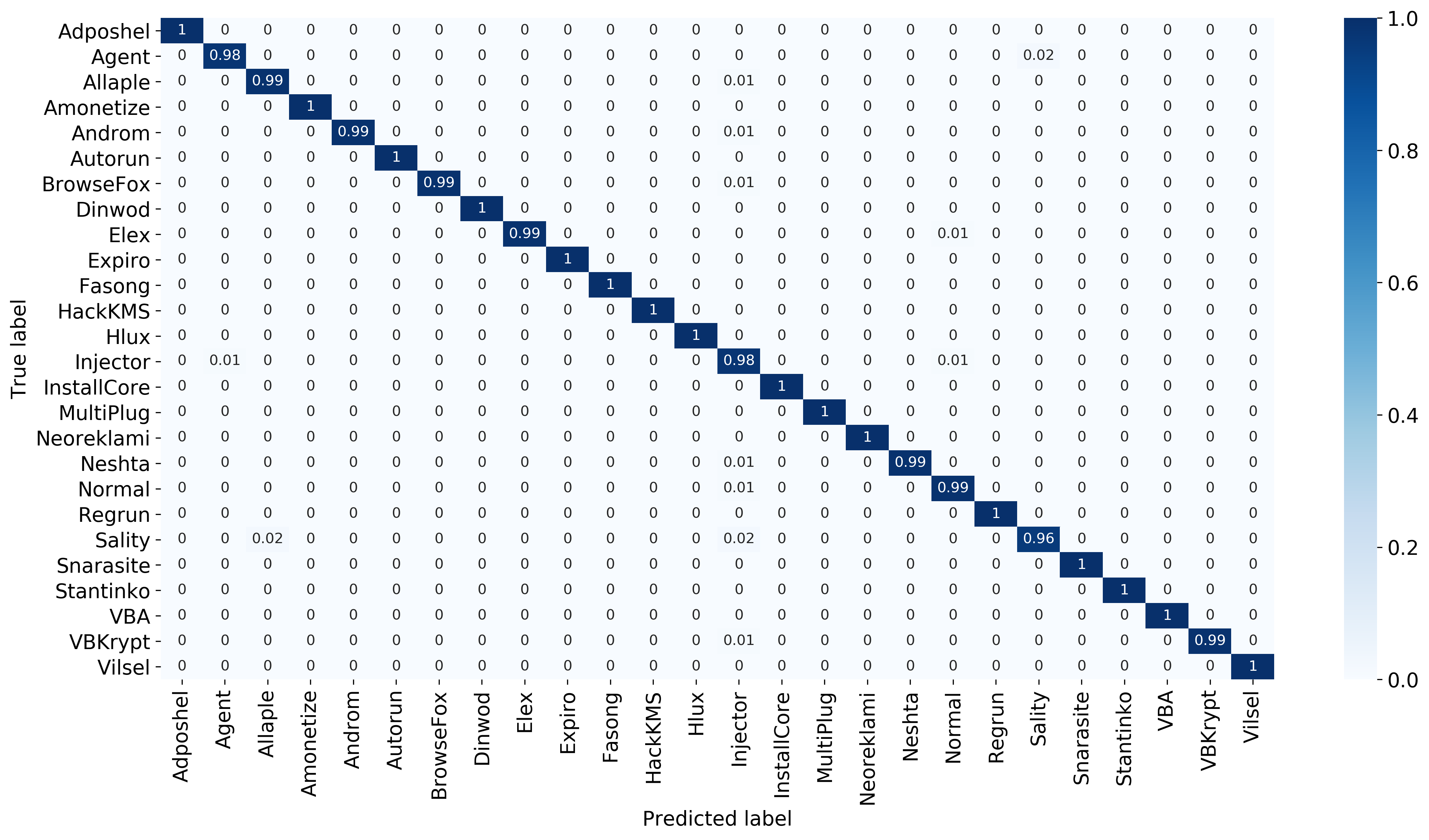}
\caption{Confusion matrix of the IoT-malware classification obtained by using FedMicro-IDA.}
\label{fig:pictfl}
\end{figure}

\begin{figure}[!h]
\centering
\includegraphics[scale=0.35]{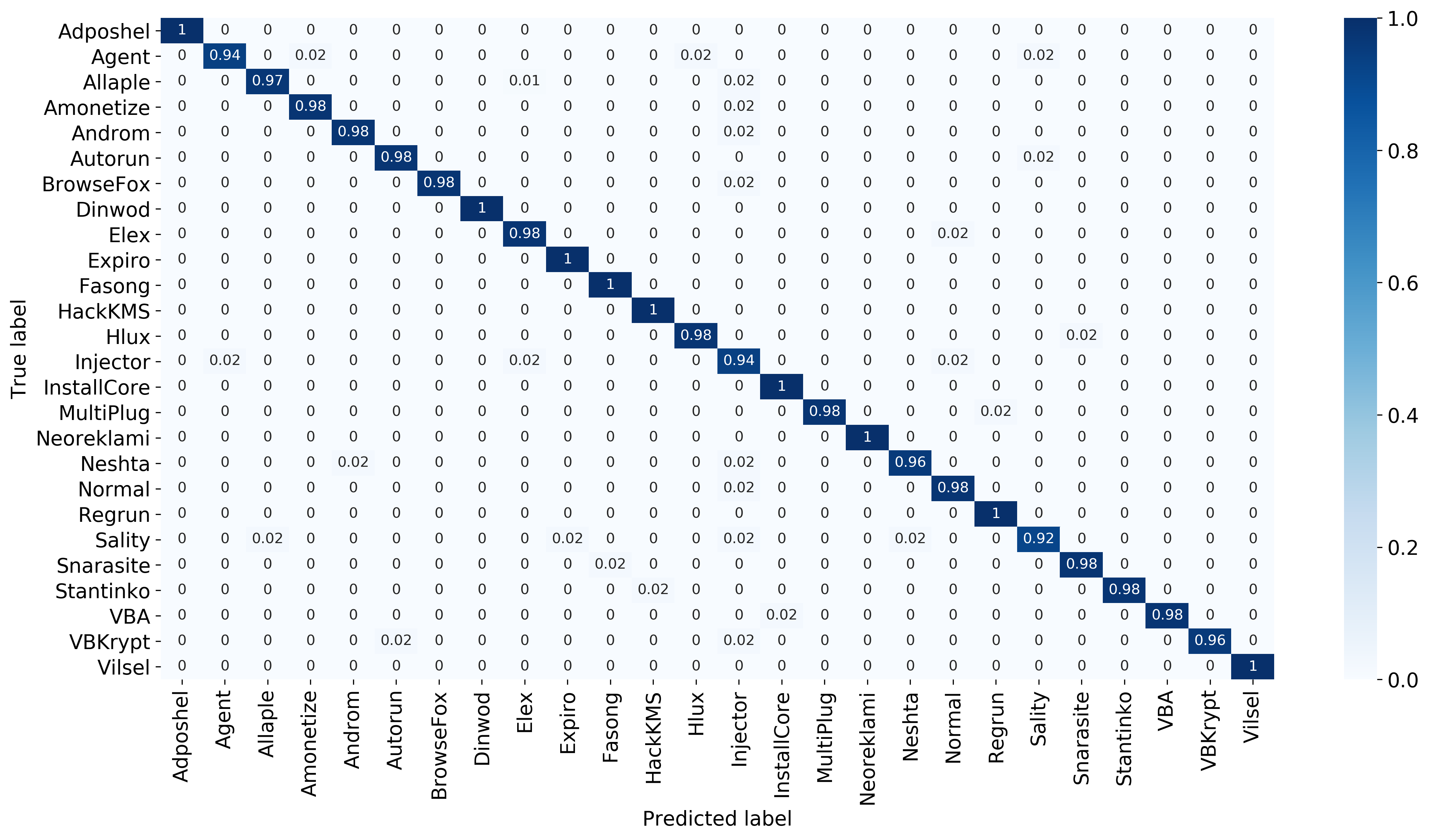}
\caption{Confusion matrix of the IoT-malware classification obtained by using the CL variant of FedMicro-ID.}
\label{fig:pictCl}
\end{figure}

\begin{figure}[!h]
\centering
\includegraphics[scale=0.5]{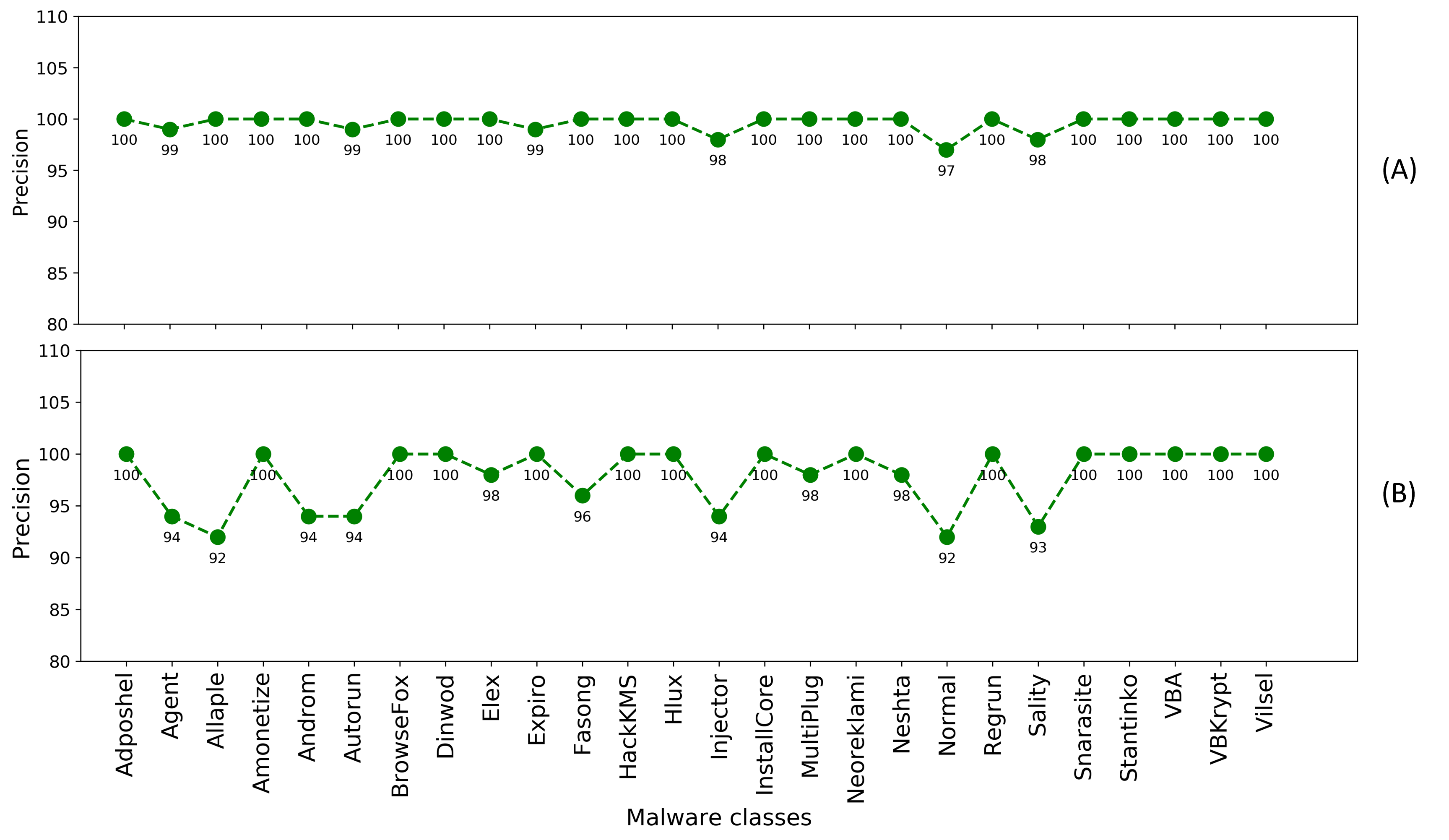}
\caption{Precision of each malware class by using: (A) the proposed FedMicro-IDA, and (B) its CL variant.}
\label{fig:precision}
\end{figure}

\begin{figure}[!h]
\centering
\includegraphics[scale=0.5]{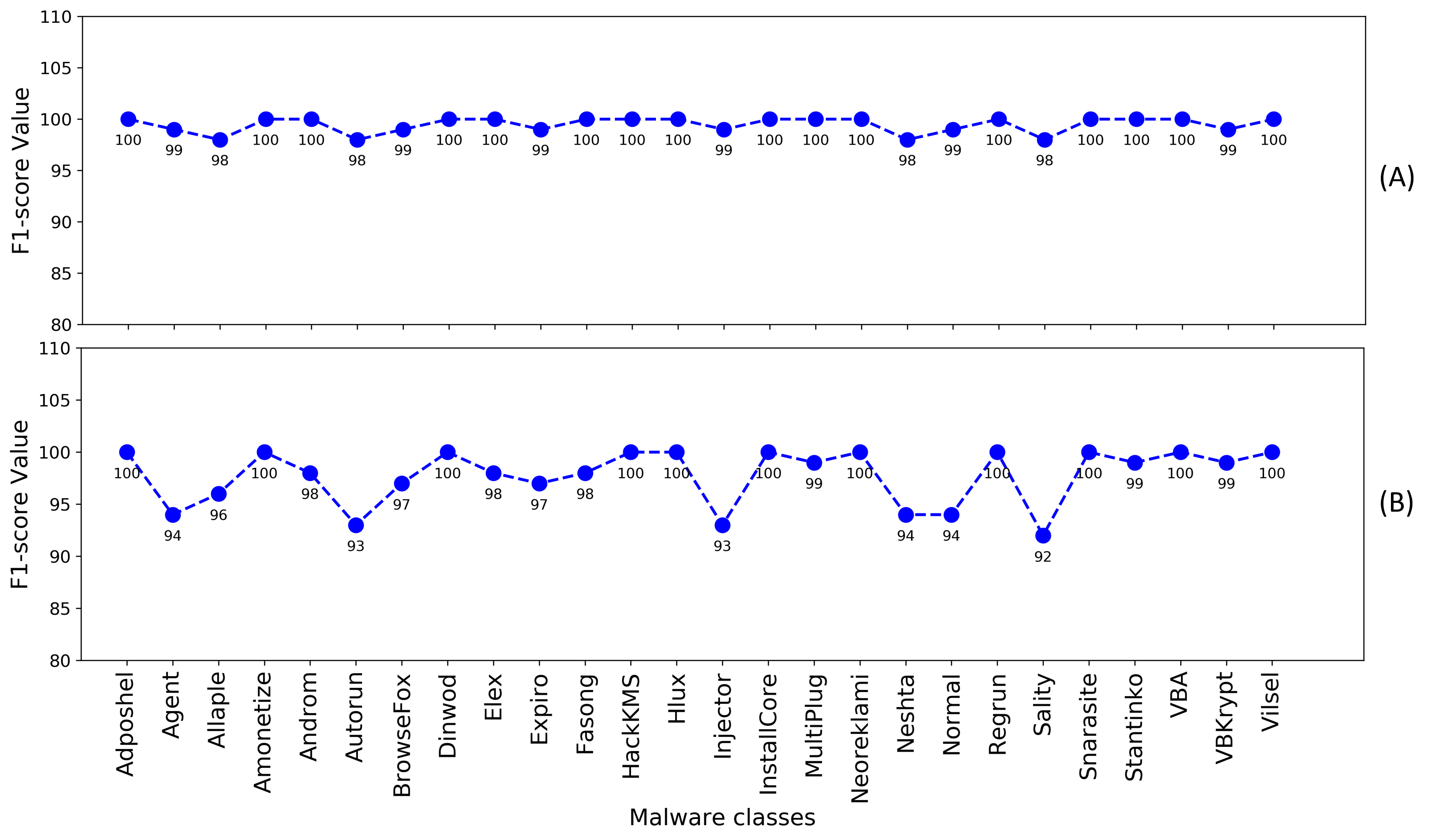}
\caption{F1-score of each malware class by using: (A) the proposed FedMicro-IDA, and (B) its CL variant.}
\label{fig:f1score}
\end{figure}

\begin{table}[h]
 \caption{Performance results of the single CNN models, the centralized Demspher-based fusion classifier, and the proposed learning in FedMicro-IDA.}
\label{tab:tab4}
\resizebox{\textwidth}{!}{%
\begin{tabular}{|c|cc|cc|cc|c|c|}
\hline
\multirow{2}{*}{\textbf{\begin{tabular}[c]{@{}c@{}}Model \textbackslash \\ Evaluation metric\end{tabular}}} &
  \multicolumn{2}{c|}{\textbf{MobileNetV2}} &
  \multicolumn{2}{c|}{\textbf{DenseNet201}} &
  \multicolumn{2}{c|}{\textbf{InceptionV3}} &
  \multicolumn{1}{c|}{\multirow{2}{*}{\textbf{\begin{tabular}[c]{@{}c@{}} CL Demspher-based \\fusion classifier \end{tabular}}}} &
  \multicolumn{1}{c|}{\multirow{2}{*}{\textbf{\begin{tabular}[c]{@{}c@{}}Proposed \\FedMicro-IDA \end{tabular}}}} 
  \\ \cline{2-7}
 &
  \multicolumn{1}{c|}{\textit{\textbf{FL}}} &
  \textit{\textbf{CL}} &
  \multicolumn{1}{c|}{\textit{\textbf{FL}}} &
  \textit{\textbf{CL}} &
  \multicolumn{1}{c|}{\textit{\textbf{FL}}} &
  \textit{\textbf{CL}} &
  \multicolumn{1}{c|}{} \\ \hline
\textbf{Accuracy(\%)}    & \multicolumn{1}{c|}{98.56}     & 93.95  & \multicolumn{1}{c|}{98.27}        & 95.12  & \multicolumn{1}{c|}{97.98}  & 93.49  & 97.15 & 99.24   \\ \hline
\textbf{Precision(\%)}   & \multicolumn{1}{c|}{98.57}     & 93.98     & \multicolumn{1}{c|}{98.35}     & 95.14  & \multicolumn{1}{c|}{97.87}     & 93.65     & 96.87 & 99.12      \\ \hline
\textbf{Recall(\%)}      & \multicolumn{1}{c|}{98.68}     & 93.84     & \multicolumn{1}{c|}{98.12}     & 95.18   & \multicolumn{1}{c|}{97.78}     & 93.23     & 97.02 & 99.36      \\ \hline
\textbf{F1-score(\%)}    & \multicolumn{1}{c|}{98.52}     & 93.81     & \multicolumn{1}{c|}{98.24}     & 95.21  & \multicolumn{1}{c|}{98.94}     & 93.21     & 97.48 & 99.45      \\ \hline
\textbf{Loss}        & \multicolumn{1}{c|}{0.1072} & 0.2939 & \multicolumn{1}{c|}{0.1251} & 0.2483 & \multicolumn{1}{c|}{0.1538} & 0.2462 & 0.1541 & 0.08471 \\ \hline
\textbf{Specificity(\%)} & \multicolumn{1}{c|}{98.78}     & 93.25    & \multicolumn{1}{c|}{98.15}      & 95.09  & \multicolumn{1}{c|}{97.74}       & 93.15    & 97.44      &  99.31\\ \hline
\textbf{MCC(\%)}         & \multicolumn{1}{c|}{98.48}     & 93.89  & \multicolumn{1}{c|}{98.18}      &   95.23    & \multicolumn{1}{c|}{97.57}       &   93.41  &  97.27     & 99.18 \\ \hline
\end{tabular}}
\end{table}
%
%
Figures \ref{fig:pictfl} and \ref{fig:pictCl} depict the confusion matrix of the proposed FedMicro-IDA and the CL Demspher-based fusion classifier, respectively. Obviously, the proposed FedMicro-IDA significantly improved recall in detecting the different types of malware. 
The plots of precision and F1-score metrics produced by FedMicro-IDA and its CL variant are shown in Figures \ref{fig:precision} and \ref{fig:f1score}. We conclude from these figures that the suggested approach yields superior results. This can be explained by the fact that the DL models are trained individually and jointly utilizing the FL approach. This allows for more effective learning by dividing the learning process over numerous clients and pooling the results to produce enhanced performance.

%
%

\begin{figure}[h]
\label{fig:scalability}
\centering
\includegraphics[scale=0.6]{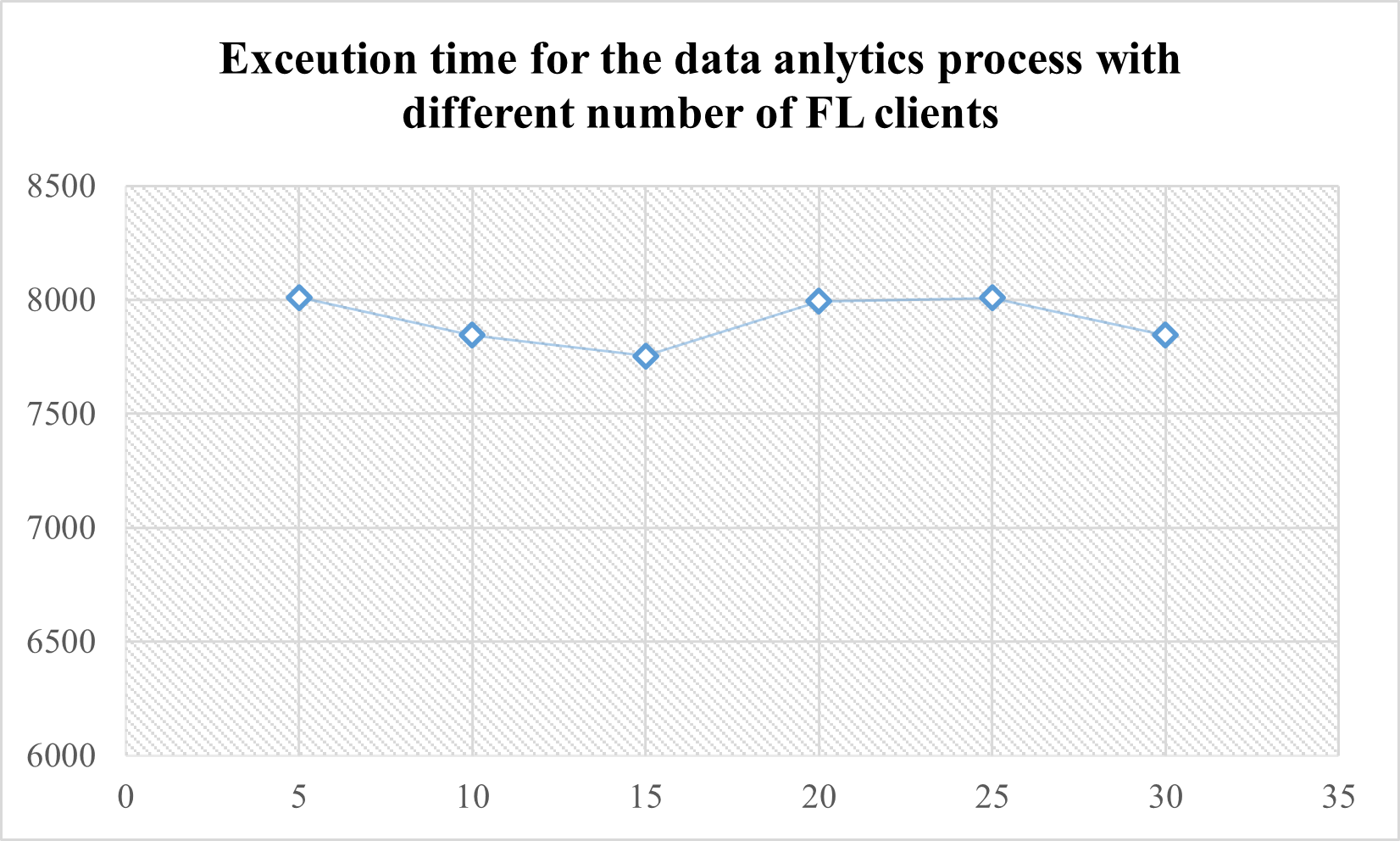}
\caption{Execution time in ms for data analytics process with a different number of FL clients.}
\label{fig:FLt}
\end{figure}

Let us now replicate the previous learning process with a larger number of data owners. Instead of decomposing the train set into 10 clients, we will breakdown it into 15, 20, 25, and 30 clients. Figure  \ref{fig:FLt} depicts the average calculation time of the learning process for the number of clients considered.
As a result, varying the number of clients in the FL process doesn't significantly worsen the overall execution time. Here are the main reasons for this conclusion:
\begin{itemize}
\item Asynchronous Communication: clients communicated with the server independently, without waiting for other clients. This can help reduce the overall execution time, as clients do not have to wait for slow or unresponsive peers. As a result, adding more clients may not necessarily increase the overall execution time since each client can operate independently;

\item Parallel Processing: the local computations for multiple clients were performed in parallel, which helped reduce the overall time required to complete a training round;

\item Sample Size: each client trained the models on a subset of its local data. Here the sample size is relatively small; adding more clients may not significantly impact the overall execution time since each client's contribution to the resulting model update was relatively small;

\item Efficient Aggregation: The server aggregated the model updates received from clients to update the global model. Here an efficient aggregation strategy, which is federated averaging, was used, so the overhead of aggregating updates from a large number of clients was minimized.

\end{itemize}

\subsubsection{Performance Analysis of Microservices}
The suggested framework has been evaluated in terms of microservices performance. For this purpose, we looked at the execution time of microservices based on their functionality. The end-to-end response time of the data analytics function ($f$) refers to the entire procedure from start to finish. It is defined by Equation \ref{eq:eqT1}:
\begin{equation}
\label{eq:eqT1}
    T_{f} = T_{Pre} + T_{Pro} + T_{Fus}
\end{equation}
The time spent in data pre-processing, data analytics, and results fusion stages are denoted as $ T_{Pre}, T_{Pro}, T_{Fus}$, respectively.

The total time cost ($T_{total}$) of data analytics function corresponds to the response time of its composed microservices and is calculated using Equations \ref{eq:eqT2} and \ref{eq:eqT3} \cite{houmani2021enabling}:
\begin{equation}
\label{eq:eqT2}
    T_{total} = \sum_{i=1}^n RT_{\mu},
\end{equation}
Where
\begin{equation}
\label{eq:eqT3}
    RT_{\mu}=Trans(M_{j}, M_{i}) + T_{Exec}
\end{equation}
$T_{Exec}$ presents the time needed for a microservice execution. $Trans(m_{j}, m_{i})$ represents the duration of data transfer from microservice $M_{j}$ to microservice $M_{i}$ and it is computed using Equation \ref{eq:eqT4}:
\begin{equation}
\label{eq:eqT4}
    Trans(M_j, M_i)  =\left\{
    \begin{split}
        \frac{size}{bw(m_j,m_n)}+\frac{size}{bw(m_n,m_i)}  \quad if r_j \in edge, r_i \in cloud \\
        0 \quad \quad \quad \quad  \quad \quad\quad \quad \quad \quad \quad \quad  r_i=r_j\\
        \frac{size}{bw(m_j,m_i)} \quad \quad \quad \quad \quad \quad \quad \quad otherwise
    \end{split}
    \right.
\end{equation}
Where $size$ is the data size transferred in Mbits, and $bw(r_j, r_n)$ is the bandwidth of the link between the data creator and data consumer in Mbits.

\begin{figure}[h]
\centering
\includegraphics[scale=0.6]{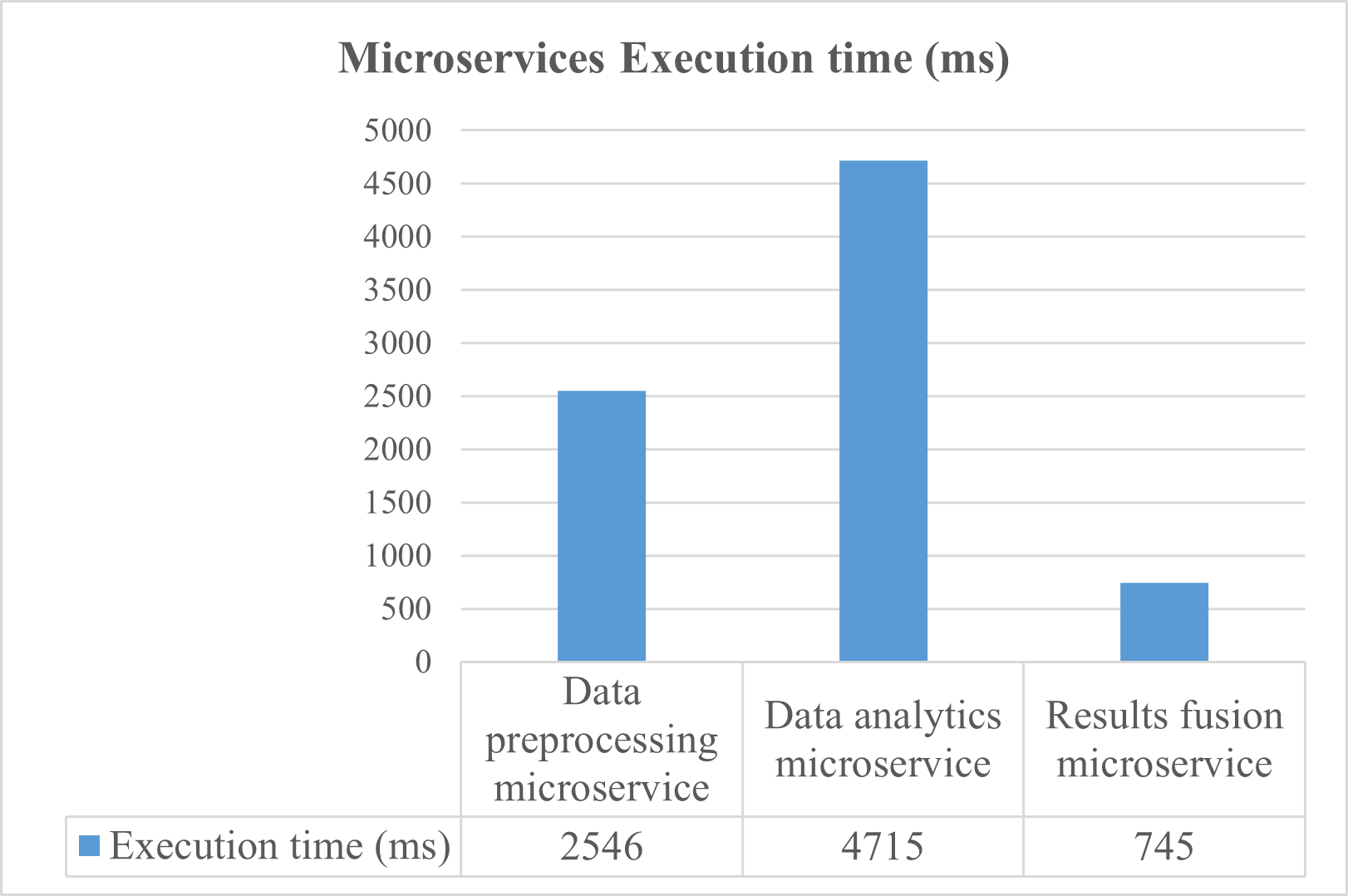}
\caption{The execution time of different microservices in the process of malware detection.}
\label{fig:ET}
\end{figure}

Figure \ref{fig:ET} depicts the execution time for each microservice in the malware detection/classification process, starting with data preparation and finishing with the fusion of the results. The data preprocessing microservice converts the recorded traffic from Portable Execution (PE) file to an RGB image, which takes approximately three seconds to complete. The data analytics function is the most time-consuming because the data is processed through three different CNN models. These models analyze the transformed image, and the classification results, expressed as probabilities for each class, are the provided outputs. Because it includes the selection of the best output based on the collected results, the fusion microservice operates significantly faster than the preceding microservices. The microservices-based architecture allows the development of smaller, specialized services that can be independently scaled. This flexibility offers faster response times and improved overall performance since the application can effectively adjust to changing loads and resource requirements. Contrarily, in a monolithic architecture, all the components of an application are tightly coupled. These components must be deployed together, making it challenging to scale individual entities to accommodate higher workloads. Consequently, this can lead to longer response times and decreased overall performance.

%
%
%
\subsubsection{Comparison with Similar Works}
To validate the proposed FedMicro-IDA, we compare the obtained results with those of previously published studies that worked on the same use case scenario and used the same dataset. As shown by the performance results in Table \ref{tab:comp}, our approach achieves the highest performance results compared to the related studies that rely on centralized ML/DL models and use monolithic architectures.
\begin{table}[h!]
\centering
\caption{Performance comparison between our approach and related approaches that use MaleVis dataset}
\label{tab:comp}
\resizebox{\textwidth}{!}{%
\begin{tabular}{|c|c|c|c|c|c|}
\hline
{ \textbf{Ref}} &
  { \textbf{Method}} &
  { \textbf{\begin{tabular}[c]{@{}c@{}}Precision(\%)\end{tabular}}} &
  { \textbf{\begin{tabular}[c]{@{}c@{}}Recall(\%)\end{tabular}}} &
  { \textbf{\begin{tabular}[c]{@{}c@{}}F1-score(\%)\end{tabular}}} &
  { \textbf{\begin{tabular}[c]{@{}c@{}}Accuracy (\%)\end{tabular}}} 
  \\ \hline
{ {\cite{bozkir2019utilization}}} &
  { \begin{tabular}[c]{@{}c@{}}TL with DenseNet201\end{tabular}} &
  { \begin{tabular}[c]{@{}c@{}}not  stated\end{tabular}} &
  { \begin{tabular}[c]{@{}c@{}}not  stated\end{tabular}} &
  { \begin{tabular}[c]{@{}c@{}}not  stated\end{tabular}} &
  { 97.48} 
  \\ \hline
 
  {{ {\cite{roseline2020intelligent}}}} &
    {{ \begin{tabular}[c]{@{}c@{}}Deep random \\ forest approach\end{tabular}}} &
    {{ 97.43}} &
    {{ 97.32}} &
    {{ 97.42}} &
    {{ 97.43}} 
  \\ \hline
{ {\cite{aslan2021new}}} &
  { \begin{tabular}[c]{@{}c@{}}Hybrid CNN \\ approach using\\AlexNet and ResNet152\end{tabular}} &
  { 97.1} &
  { 94.9} &
  { 94.5} &
  { 96.6} 
  \\ \hline
{ {\cite{wong2021vision}}} &
  { \begin{tabular}[c]{@{}c@{}}TL with ShuffleNet\\ and DenseNet201\end{tabular}} &
  { {99.80}} &
  { 95.61} &
  { 95.37} &
  { 95.01} 
  \\
  \hline
{ {\cite{hemalatha2021efficient}}} &
  { \begin{tabular}[c]{@{}c@{}}Pretrained \\DenseNet model\end{tabular}} &
  { 98.56} &
  { 97.74} &
  { 98.15} &
  { 98.21} 
 \\ \hline
{ {\begin{tabular}[c]{@{}c@{}}\cite{atitallah2022novel}\end{tabular}}} &
  { {\begin{tabular}[c]{@{}c@{}}Random forest-based\\  voting classifier\end{tabular}}} &
  { 98.74} &
  { {98.67}} &
  { { 98.70}} &
  { {98.98}} 
 \\ \hline
 { {\begin{tabular}[c]{@{}c@{}} FedMicro-IDA\end{tabular} }} &
  { \begin{tabular}[c]{@{}c@{}} FL and microservices\\based framework for\\ IoT
data analytics \end{tabular}} 
  &
  {99.12} &
  {99.36} &
  {99.45} &
  {99.24} 
  \\
  \hline
\end{tabular}}
\end{table}

The present research demonstrates the usefulness of using distributed learning methodology through FL, in addition to TL, to build more accurate models that guarantee efficient data analytical tasks.
Moreover, by implementing each task in the learning process as a microservice, the FedMicro-IDA improves data analytics development, deployment, and maintainability. The utilization of tiny and independent services speeds up the execution of the various steps in the analytics process, improves modularity and scalability, and reinforces fault isolation.
\section{Conclusion}
In this paper, we proposed the FedMicro-IDA approach combining FL, TL, and ensemble learning to support data analytics in microservices-based IoT applications. 
The suggested approach produced learning models with high performance through knowledge transfer and ensured the privacy and security of clients' data through local training. Experiments on malware detection and classification use case have proved the FedMicro-IDA usefulness and effectiveness. 
In future work, we aim to implement our approach in real-world distributed environments. This will allow us to evaluate and validate its performance, scalability, and effectiveness in real-world applications/scenarios.

To enhance our proposed framework, a potential avenue for future research lies in exploring novel methods for optimizing communication and collaboration between different microservices. This could involve developing more efficient and secure ways for microservices to share data and insights, such as through the use of blockchain or other distributed ledger technologies.
Another direction could be to investigate new ways to optimize the deployment and management of microservices across distributed and heterogeneous environments. This could involve developing new containerization and orchestration technologies that can more effectively manage and coordinate the deployment of microservices across different IoT devices and cloud platforms.
\section*{Acknowledgment}
The authors would like to thank Prince Sultan University for their support.

%
%

\printbibliography

@article{atitallah2020leveraging,
  title={Leveraging Deep Learning and IoT big data analytics to support the smart cities development: Review and future directions},
  author={Ben Atitallah, Safa and Driss, Maha and Boulila, Wadii and Ben Gh{\'e}zala, Henda},
  journal={Computer Science Review},
  volume={38},
  pages={100303},
  year={2020},
  publisher={Elsevier}
}

@article{hajjaji2021big,
  title={Big data and IoT-based applications in smart environments: A systematic review},
  author={Hajjaji, Yosra and Boulila, Wadii and Farah, Imed Riadh and Romdhani, Imed and Hussain, Amir},
  journal={Computer Science Review},
  volume={39},
  pages={100318},
  year={2021},
  publisher={Elsevier}
}

@article{driss2022federated,
  title={A federated learning framework for cyberattack detection in vehicular sensor networks},
  author={Driss, Maha and Almomani, Iman and Ahmad, Jawad and others},
  journal={Complex \& Intelligent Systems},
  pages={1--15},
  year={2022},
  publisher={Springer}
}

@article{mezni2022smartwater,
  title={Smartwater: A service-oriented and sensor cloud-based framework for smart monitoring of water environments},
  author={Mezni, Haithem and Driss, Maha and Boulila, Wadii and Atitallah, Safa Ben and Sellami, Mokhtar and Alharbi, Nouf},
  journal={Remote Sensing},
  volume={14},
  number={4},
  pages={922},
  year={2022},
  publisher={MDPI}
}

@article{saleem2019data,
  title={Data analytics in the Internet of Things: a survey},
  author={Saleem, Tausifa Jan and Chishti, Mohammad Ahsan},
  journal={Scalable Computing: Practice and Experience},
  volume={20},
  number={4},
  pages={607--630},
  year={2019}
}

@article{nguyen2021federated,
  title={Federated learning for internet of things: A comprehensive survey},
  author={Nguyen, Dinh C and Ding, Ming and Pathirana, Pubudu N and Seneviratne, Aruna and Li, Jun and Poor, H Vincent},
  journal={IEEE Communications Surveys \& Tutorials},
  year={2021},
  publisher={IEEE}
}

@article{khan2021federated,
  title={Federated learning for internet of things: Recent advances, taxonomy, and open challenges},
  author={Khan, Latif U and Saad, Walid and Han, Zhu and Hossain, Ekram and Hong, Choong Seon},
  journal={IEEE Communications Surveys \& Tutorials},
  year={2021},
  publisher={IEEE}
}

@article{al2022ai,
  title={AI-enabled Secure Microservices in Edge Computing: Opportunities and Challenges},
  author={Al-Doghman, Firas and Moustafa, Nour and Khalil, Ibrahim and Tari, Zahir and Zomaya, Albert},
  journal={IEEE Transactions on Services Computing},
  year={2022},
  publisher={IEEE}
}

@misc{Anaconda,
  title = {Anaconda},
  howpublished = {\url{https://www.anaconda.com/}},
  note = {Accessed: January 19, 2023}
}

@misc{kubernetes,
  title = {Docker Orchestration},
  howpublished = {\url{https://docs.docker.com/get-started/orchestration/}},
  note = {Accessed: January 19, 2023}
}

@misc{Jupyter,
  title = {Jupyter: Free software, open standards, and web services for interactive computing across all programming languages},
  howpublished = {\url{https://jupyter.org/}},
  note = {Accessed: January 19, 2023}
}

@misc{MaleVis,
  title = {MaleVis dataset},
  howpublished = {\url{https://web.cs.hacettepe.edu.tr/~selman/malevis/}},
  note = {Accessed: January 19, 2023}
}

@misc{Python,
  title = {Python programming language},
  howpublished = {\url{https://www.python.org/}},
  note = {Accessed: January 19, 2023}
}

@misc{tff,
  title = {Federated transfer learning},
  howpublished = {\url{https://www.tensorflow.org/federated/get_started}},
  note = {Accessed: January 19, 2023}
}

@misc{tf,
  title = {Create production-grade machine learning models with TensorFlow},
  howpublished = {\url{https://www.tensorflow.org/overview}},
  note = {Accessed: January 19, 2023}
}

@article{jamil2021intelligent,
  title={Intelligent microservice based on blockchain for healthcare applications},
  author={Jamil, Faisal and Qayyum, Faiza and Alhelaly, Soha and Javed, Farjeel and Muthanna, Ammar},
  journal={CMC-Comput. Mater. Contin},
  volume={69},
  pages={2513--2530},
  year={2021}
}

@inproceedings{dineva2020architectural,
  title={Architectural ML framework for IoT services delivery based on microservices},
  author={Dineva, Kristina and Atanasova, Tatiana},
  booktitle={International Conference on Distributed Computer and Communication Networks},
  pages={698--711},
  year={2020},
  organization={Springer}
}

@article{ortiz2019real,
  title={Real-time context-aware microservice architecture for predictive analytics and smart decision-making},
  author={Ortiz, Guadalupe and Caravaca, Jos{\'e} Antonio and Garc{\'\i}a-de-Prado, Alfonso and Boubeta-Puig, Juan and others},
  journal={IEEE Access},
  volume={7},
  pages={183177--183194},
  year={2019},
  publisher={IEEE}
}

@article{ali2018design,
  title={Design methodology of microservices to support predictive analytics for IoT applications},
  author={Ali, Sajjad and Jarwar, Muhammad Aslam and Chong, Ilyoung},
  journal={Sensors},
  volume={18},
  number={12},
  pages={4226},
  year={2018},
  publisher={Multidisciplinary Digital Publishing Institute}
}

@article{zhuang2020comprehensive,
  title={A comprehensive survey on transfer learning},
  author={Zhuang, Fuzhen and Qi, Zhiyuan and Duan, Keyu and Xi, Dongbo and Zhu, Yongchun and Zhu, Hengshu and Xiong, Hui and He, Qing},
  journal={Proceedings of the IEEE},
  volume={109},
  number={1},
  pages={43--76},
  year={2020},
  publisher={IEEE}
}

@article{atitallah2022novel,
  title={A Novel Detection and Multi-Classification Approach for IoT-Malware Using Random Forest Voting of Fine-Tuning Convolutional Neural Networks},
  author={Atitallah, Safa Ben and Driss, Maha and Almomani, Iman},
  journal={Sensors},
  volume={22},
  number={11},
  pages={4302},
  year={2022},
  publisher={MDPI}
}

@article{wu2020exploiting,
  title={Exploiting transfer learning for emotion recognition under cloud-edge-client collaborations},
  author={Wu, Dapeng and Han, Xiaojuan and Yang, Zhigang and Wang, Ruyan},
  journal={IEEE Journal on Selected Areas in Communications},
  volume={39},
  number={2},
  pages={479--490},
  year={2020},
  publisher={IEEE}
}

@article{ohata2020automatic,
  title={Automatic detection of COVID-19 infection using chest X-ray images through transfer learning},
  author={Ohata, Elene Firmeza and Bezerra, Gabriel Maia and das Chagas, Joao Victor Souza and Neto, Alo{\'\i}sio Vieira Lira and Albuquerque, Adriano Bessa and de Albuquerque, Victor Hugo C and Reboucas Filho, Pedro Pedrosa},
  journal={IEEE/CAA Journal of Automatica Sinica},
  volume={8},
  number={1},
  pages={239--248},
  year={2020},
  publisher={IEEE}
}

@inproceedings{sandler2018mobilenetv2,
  title={Mobilenetv2: Inverted residuals and linear bottlenecks},
  author={Sandler, Mark and Howard, Andrew and Zhu, Menglong and Zhmoginov, Andrey and Chen, Liang-Chieh},
  booktitle={Proceedings of the IEEE conference on computer vision and pattern recognition},
  pages={4510--4520},
  year={2018}
}

@inproceedings{huang2017densely,
  title={Densely connected convolutional networks},
  author={Huang, Gao and Liu, Zhuang and Van Der Maaten, Laurens and Weinberger, Kilian Q},
  booktitle={Proceedings of the IEEE conference on computer vision and pattern recognition},
  pages={4700--4708},
  year={2017}
}

@article{chen2020fedhealth,
  title={Fedhealth: A federated transfer learning framework for wearable healthcare},
  author={Chen, Yiqiang and Qin, Xin and Wang, Jindong and Yu, Chaohui and Gao, Wen},
  journal={IEEE Intelligent Systems},
  volume={35},
  number={4},
  pages={83--93},
  year={2020},
  publisher={IEEE}
}

@article{abdel2021federated,
  title={Federated threat-hunting approach for microservice-based industrial cyber-physical system},
  author={Abdel-Basset, Mohamed and Hawash, Hossam and Sallam, Karam},
  journal={IEEE Transactions on Industrial Informatics},
  volume={18},
  number={3},
  pages={1905--1917},
  year={2021},
  publisher={IEEE}
}

@article{he2020group,
  title={Group knowledge transfer: Federated learning of large cnns at the edge},
  author={He, Chaoyang and Annavaram, Murali and Avestimehr, Salman},
  journal={Advances in Neural Information Processing Systems},
  volume={33},
  pages={14068--14080},
  year={2020}
}

@article{shi2022deep,
  title={Deep Federated Adaptation: An Adaptative Residential Load Forecasting Approach with Federated Learning},
  author={Shi, Yuan and Xu, Xianze},
  journal={Sensors},
  volume={22},
  number={9},
  pages={3264},
  year={2022},
  publisher={MDPI}
}

@article{kevin2021federated,
  title={Federated transfer learning based cross-domain prediction for smart manufacturing},
  author={Kevin, I and Wang, Kai and Zhou, Xiaokang and Liang, Wei and Yan, Zheng and She, Jinhua},
  journal={IEEE Transactions on Industrial Informatics},
  volume={18},
  number={6},
  pages={4088--4096},
  year={2021},
  publisher={IEEE}
}

@article{driss2021microservices,
  title={Microservices in IoT security: current solutions, research challenges, and future directions},
  author={Driss, Maha and Hasan, Daniah and Boulila, Wadii and Ahmad, Jawad},
  journal={Procedia Computer Science},
  volume={192},
  pages={2385--2395},
  year={2021},
  publisher={Elsevier}
}

@inproceedings{hasan2021sublmume,
  title={SUBL$\mu$ME: Secure Blockchain as a Service and Microservices-based Framework for IoT Environments},
  author={Hasan, Daniah and Driss, Maha},
  booktitle={2021 IEEE/ACS 18th International Conference on Computer Systems and Applications (AICCSA)},
  pages={1--9},
  year={2021},
  organization={IEEE}
}

@article{driss2022ws,
  title={WS-ADVISING: a Reusable and reconfigurable microservices-based platform for effective academic advising},
  author={Driss, Maha},
  journal={Journal of Ambient Intelligence and Humanized Computing},
  volume={13},
  number={1},
  pages={283--294},
  year={2022},
  publisher={Springer}
}

@article{driss2022req,
  title={Req-WSComposer: a novel platform for requirements-driven composition of semantic web services},
  author={Driss, Maha and Ben Atitallah, Safa and Albalawi, Amal and Boulila, Wadii},
  journal={Journal of Ambient Intelligence and Humanized Computing},
  volume={13},
  number={2},
  pages={849--865},
  year={2022},
  publisher={Springer}
}

@article{abreha2022federated,
  title={Federated learning in edge computing: a systematic survey},
  author={Abreha, Haftay Gebreslasie and Hayajneh, Mohammad and Serhani, Mohamed Adel},
  journal={Sensors},
  volume={22},
  number={2},
  pages={450},
  year={2022},
  publisher={MDPI}
}

@article{li2020review,
  title={A review of applications in federated learning},
  author={Li, Li and Fan, Yuxi and Tse, Mike and Lin, Kuo-Yi},
  journal={Computers \& Industrial Engineering},
  volume={149},
  pages={106854},
  year={2020},
  publisher=
  {Elsevier}
}

@article{liu2020secure,
  title={A secure federated transfer learning framework},
  author={Liu, Yang and Kang, Yan and Xing, Chaoping and Chen, Tianjian and Yang, Qiang},
  journal={IEEE Intelligent Systems},
  volume={35},
  number={4},
  pages={70--82},
  year={2020},
  publisher={IEEE}
}

@article{zhang2022federated,
  title={Federated transfer learning for disaster classification in social computing networks},
  author={Zhang, Zehui and He, Ningxin and Li, Donagyu and Gao, Hang and Gao, Tiegang and Zhou, Chuan},
  journal={Journal of Safety Science and Resilience},
  volume={3},
  number={1},
  pages={15--23},
  year={2022},
  publisher={Elsevier}
}

@article{atitallah2022microservices,
  title={Microservices for Data Analytics in IoT Applications: Current Solutions, Open Challenges, and Future Research Directions},
  author={Ben Atitallah, Safa and Driss, Maha and Ben Ghzela, Henda},
  journal={Procedia Computer Science},
  volume={207},
  pages={3938--3947},
  year={2022},
  publisher={Elsevier}
}

@misc{docker,
  title = {Docker documentation},
  howpublished = {\url{https://docs.docker.com/}},
  note = {Accessed: November 6, 2022}
}

@article{ben2022randomly,
  title={Randomly initialized convolutional neural network for the recognition of COVID-19 using X-ray images},
  author={Ben Atitallah, Safa and Driss, Maha and Boulila, Wadii and Ben Ghezala, Henda},
  journal={International journal of imaging systems and technology},
  volume={32},
  number={1},
  pages={55--73},
  year={2022},
  publisher={Wiley Online Library}
}

@article{ben2022fusion,
  title={Fusion of convolutional neural networks based on Dempster--Shafer theory for automatic pneumonia detection from chest X-ray images},
  author={Ben Atitallah, Safa and Driss, Maha and Boulila, Wadii and Koubaa, Anis and Ben Ghezala, Henda},
  journal={International Journal of Imaging Systems and Technology},
  volume={32},
  number={2},
  pages={658--672},
  year={2022},
  publisher={Wiley Online Library}
}

@inproceedings{tan2018survey,
  title={A survey on deep transfer learning},
  author={Tan, Chuanqi and Sun, Fuchun and Kong, Tao and Zhang, Wenchang and Yang, Chao and Liu, Chunfang},
  booktitle={International conference on artificial neural networks},
  pages={270--279},
  year={2018},
  organization={Springer}
}

@article{champaneria2023microservices,
  title={Microservices in IoT Middleware Architectures: Architecture, Trends, and Challenges},
  author={Champaneria, Tushar and Jardosh, Sunil and Makwana, Ashwin},
  journal={IOT with Smart Systems},
  pages={381--395},
  year={2023},
  publisher={Springer}
}

@article{surianarayanan2019essentials,
  title={Essentials of Microservices Architecture: Paradigms, Applications, and Techniques},
  author={Surianarayanan, Chellammal and Ganapathy, Gopinath and Pethuru, Raj},
  year={2019},
  publisher={Taylor \& Francis}
}

@article{bucchiarone2020microservices,
  title={Microservices},
  author={Bucchiarone, Antonio and Dragoni, Nicola and Dustdar, Schahram and Lago, Patricia and Mazzara, Manuel and Rivera, Victor and Sadovykh, Andrey},
  journal={Science and Engineering. Springer},
  year={2020},
  publisher={Springer}
}

@article{shaik2022fedstack,
  title={FedStack: Personalized activity monitoring using stacked federated learning},
  author={Shaik, Thanveer and Tao, Xiaohui and Higgins, Niall and Gururajan, Raj and Li, Yuefeng and Zhou, Xujuan and Acharya, U Rajendra},
  journal={Knowledge-Based Systems},
  volume={257},
  pages={109929},
  year={2022},
  publisher={Elsevier}
}

@article{su2021secure,
  title={Secure and efficient federated learning for smart grid with edge-cloud collaboration},
  author={Su, Zhou and Wang, Yuntao and Luan, Tom H and Zhang, Ning and Li, Feng and Chen, Tao and Cao, Hui},
  journal={IEEE Transactions on Industrial Informatics},
  volume={18},
  number={2},
  pages={1333--1344},
  year={2021},
  publisher={IEEE}
}

@inproceedings{houmani2021enabling,
  title={Enabling microservices management for Deep Learning applications across the Edge-Cloud Continuum},
  author={Houmani, Zeina and Balouek-Thomert, Daniel and Caron, Eddy and Parashar, Manish},
  booktitle={2021 IEEE 33rd International Symposium on Computer Architecture and High Performance Computing (SBAC-PAD)},
  pages={137--146},
  year={2021},
  organization={IEEE}
}

@article{roy2021micro,
  title={Micro-Safe: Microservices-and Deep Learning-Based Safety-as-a-Service Architecture for 6G-Enabled Intelligent Transportation System},
  author={Roy, Chandana and Saha, Ruelia and Misra, Sudip and Dev, Kapal},
  journal={IEEE Transactions on Intelligent Transportation Systems},
  year={2021},
  publisher={IEEE}
}

@article{nikolakis2020microservice,
  title={A microservice architecture for predictive analytics in manufacturing},
  author={Nikolakis, N and Marguglio, A and Veneziano, G and Greco, P and Panicucci, S and Cerquitelli, T and Macii, E and Andolina, S and Alexopoulos, K},
  journal={Procedia Manufacturing},
  volume={51},
  pages={1091--1097},
  year={2020},
  publisher={Elsevier}
}

@inproceedings{vresk2016architecture,
  title={Architecture of an interoperable IoT platform based on microservices},
  author={Vresk, Tomislav and {\v{C}}avrak, Igor},
  booktitle={2016 39th International Convention on Information and Communication Technology, Electronics and Microelectronics (MIPRO)},
  pages={1196--1201},
  year={2016},
  organization={IEEE}
}

@article{attota2021ensemble,
  title={An ensemble multi-view federated learning intrusion detection for iot},
  author={Attota, Dinesh Chowdary and Mothukuri, Viraaji and Parizi, Reza M and Pouriyeh, Seyedamin},
  journal={IEEE Access},
  volume={9},
  pages={117734--117745},
  year={2021},
  publisher={IEEE}
}

@inproceedings{szegedy2015going,
  title={Going deeper with convolutions},
  author={Szegedy, Christian and Liu, Wei and Jia, Yangqing and Sermanet, Pierre and Reed, Scott and Anguelov, Dragomir and Erhan, Dumitru and Vanhoucke, Vincent and Rabinovich, Andrew},
  booktitle={Proceedings of the IEEE conference on computer vision and pattern recognition},
  pages={1--9},
  year={2015}
}

@inproceedings{mcmahan2017communication,
  title={Communication-efficient learning of deep networks from decentralized data},
  author={McMahan, Brendan and Moore, Eider and Ramage, Daniel and Hampson, Seth and y Arcas, Blaise Aguera},
  booktitle={Artificial intelligence and statistics},
  pages={1273--1282},
  year={2017},
  organization={PMLR}
}

@article{denoeux2000neural,
  title={A neural network classifier based on Dempster-Shafer theory},
  author={Denoeux, Thierry},
  journal={IEEE Transactions on Systems, Man, and Cybernetics-Part A: Systems and Humans},
  volume={30},
  number={2},
  pages={131--150},
  year={2000},
  publisher={IEEE}
}

@article{sentz2002combination,
  title={Combination of evidence in Dempster-Shafer theory},
  author={Sentz, Kari and Ferson, Scott},
  year={2002},
  publisher={Sandia National Lab.(SNL-NM), Albuquerque, NM (United States); Sandia~…}
}

@book{smith2017docker,
  title={Docker Orchestration},
  author={Smith, Randall},
  year={2017},
  publisher={Packt Publishing Ltd}
}

@inproceedings{bozkir2019utilization,
  title={Utilization and comparision of convolutional neural networks in malware recognition},
  author={Bozkir, Ahmet Selman and Cankaya, Ahmet Ogulcan and Aydos, Murat},
  booktitle={2019 27th Signal Processing and Communications Applications Conference (SIU)},
  pages={1--4},
  year={2019},
  organization={IEEE}
}

@article{roseline2020intelligent,
  title={Intelligent vision-based malware detection and classification using deep random forest paradigm},
  author={Roseline, S Abijah and Geetha, S and Kadry, Seifedine and Nam, Yunyoung},
  journal={IEEE Access},
  volume={8},
  pages={206303--206324},
  year={2020},
  publisher={IEEE}
}

@article{aslan2021new,
  title={A new malware classification framework based on deep learning algorithms},
  author={Aslan, {\"O}mer and Yilmaz, Abdullah Asim},
  journal={Ieee Access},
  volume={9},
  pages={87936--87951},
  year={2021},
  publisher={IEEE}
}

@article{wong2021vision,
  title={Vision-Based Malware Detection: A Transfer Learning Approach Using Optimal ECOC-SVM Configuration},
  author={Wong, WK and Juwono, Filbert H and Apriono, Catur},
  journal={IEEE Access},
  volume={9},
  pages={159262--159270},
  year={2021},
  publisher={IEEE}
}

@article{hemalatha2021efficient,
  title={An efficient densenet-based deep learning model for malware detection},
  author={Hemalatha, Jeyaprakash and Roseline, S Abijah and Geetha, Subbiah and Kadry, Seifedine and Dama{\v{s}}evi{\v{c}}ius, Robertas},
  journal={Entropy},
  volume={23},
  number={3},
  pages={344},
  year={2021},
  publisher={Multidisciplinary Digital Publishing Institute}
}

@misc{keras,
  title = {Keras Library},
  howpublished = {\url{https://keras.io/}},
  note = {Accessed: May 19, 2023}
}

@misc{tensorflow,
  title = {Tensorflow Library},
  howpublished = {https://www.tensorflow.org/}}

@misc{ftensorflow,
  title = {TensorFlow Federated: Machine Learning on Decentralized Data},
  howpublished = {https://www.tensorflow.org/federated}}

@article{bibi2022deep,
  title={Deep AI-powered Cyber Threat Analysis in IIoT},
  author={Bibi, Iram and Akhunzada, Adnan and Kumar, Neeraj},
  journal={IEEE Internet of Things Journal},
  year={2022},
  publisher={IEEE}
}
\end{document}